\documentclass[12pt,letter]{article}
\usepackage{graphicx, epsfig, color, amssymb}
\def\eslt{E_T^{\rm miss}}
\def\to{\rightarrow}

\def\bi{\begin{itemize}}
\def\ei{\end{itemize}}
\def\te{\tilde e}

\def\ta{\tilde a}

\def\tu{\tilde u}

\def\tb{\tilde b}
\def\tf{\tilde f}

\def\tst{\tilde t}
\def\ttau{\tilde \tau}

\def\tg{\tilde g}
\def\tnu{\tilde\nu}

\def\tq{\tilde q}
\def\tw{\widetilde W}
\def\tz{\widetilde Z}
\def\alt{\stackrel{<}{\sim}}
\def\agt{\stackrel{>}{\sim}}
\def\be{\begin{equation}}  
\def\ee{\end{equation}}  

\newcommand\prd[3]{{\it Phys.\ Rev.\ }{\bf D #1} (#2) #3}
\newcommand\prep[3]{{\it Phys.\ Rept.\ }{\bf #1} (#2) #3}
\newcommand\prl[3]{{\it Phys.\ Rev.\ Lett.\ }{\bf #1} (#2) #3}
\newcommand\plb[3]{{\it Phys.\ Lett.\ }{\bf B #1} (#2) #3}

\newcommand\jhep[3]{{\it J. High Energy Phys.\ }{\bf #1} (#2) #3}
\newcommand\app[3]{{\it Astropart.\ Phys.\ }{\bf #1} (#2) #3}

\newcommand\npb[3]{{\it Nucl.\ Phys.\ }{\bf B #1} (#2) #3}
\newcommand\epjc[3]{{\it Eur.\ Phys.\ J. }{\bf C #1} (#2) #3}

\newcommand\jcap[3]{{\it JCAP}{\bf #1} (#2) #3}
\newcommand{\hepph}[1]{hep-ph/#1}

\begin{document}
\begin{titlepage}

\begin{flushright}
LPSC10105\\
UH-511-1152-10
\end{flushright}

\vspace*{0.5cm}
\begin{center}
{\Large \bf 
Effective Supersymmetry at the LHC\\
}
\vspace{1.2cm} \renewcommand{\thefootnote}{\fnsymbol{footnote}}
{\large Howard Baer$^{1}$\footnote[1]{Email: baer@nhn.ou.edu },
Sabine Kraml$^{2}$\footnote[2]{Email: sabine.kraml@lpsc.in2p3.fr},
Andre Lessa$^{1}$\footnote[3]{Email: lessa@nhn.ou.edu},\\
Sezen Sekmen$^3$\footnote[4]{Email: sezen.sekmen@cern.ch}
and Xerxes Tata$^{4,5}$\footnote[5]{Email: tata@phys.hawaii.edu }} \\
\vspace{1.2cm} \renewcommand{\thefootnote}{\arabic{footnote}}
{\it 
$^1$Dept.\ of Physics and Astronomy,
University of Oklahoma, Norman, OK 73019, USA \\
$^2$Laboratoire de Physique Subatomique et de Cosmologie, UJF Grenoble 1, 
CNRS/IN2P3, INPG, 53 Avenue des Martyrs, F-38026 Grenoble, France\\
$^3$Dept.\ of Physics, Florida State University, Tallahassee, FL 32306, USA\\
$^4$Dept.\ of Physics and Astronomy, University of Hawaii, Honolulu, HI 96822,USA\\
$^5$Dept.\ of Physics, University of Wisconsin, Madison, WI 53706, USA
}

\end{center}

\vspace{0.5cm}
\begin{abstract}
\noindent 
We investigate the phenomenology of Effective Supersymmetry (ESUSY)
models wherein electroweak gauginos and third generation scalars
have masses up to about 1~TeV while first and second generation scalars 
lie in the multi-TeV range.  
Such models ameliorate the SUSY flavor and CP problems
via a decoupling solution, while at the same time maintaining
naturalness. In our analysis, we assume independent GUT scale mass
parameters for third and first/second generation scalars and for the
Higgs scalars, in addition to $m_{1/2}$, $\tan\beta$ and $A_0$, and
require radiative electroweak symmetry breaking as usual. 
We analyse the parameter space which is consistent with current 
constraints, by means of a Markov Chain Monte Carlo scan. 
The lightest MSSM particle (LMP) is mostly, but not always the lightest
neutralino, and moreover, the thermal relic density of the neutralino
LMP is frequently very large. These models may phenomenologically be
perfectly viable if the LMP before nucleosynthesis decays into the axino
plus SM particles. Dark matter is then an axion/axino mixture. At the
LHC, the most important production mechanisms are gluino production (for
$m_{1/2} \alt 700$~GeV) and third generation squark production, while
SUSY events rich in $b$-jets are the hallmark of the ESUSY scenario. 
We present a set of ESUSY benchmark points with characteristic 
features and discuss their LHC phenomenology.\\

\noindent PACS numbers: 14.80.Ly, 12.60.Jv, 11.30.Pb

\end{abstract}

\end{titlepage}

\section{Introduction}
\label{sec:intro}

Particle physics models that include weak scale softly broken
supersymmetry (SUSY) are especially compelling in that they stabilize
the weak scale, even in the presence of new physics at much higher
energy scales, such as $M_{\rm Planck}$ or $M_{\rm GUT}$. 
The simplest of such models, the Minimal Supersymmetric Standard
Model (MSSM) \cite{dg,sakai} enjoys some compelling indirect
experimental support in that the measured values of the gauge couplings
at energy scale $Q=M_Z$, when run to high energies via renormalization
group evolution (RGE), nearly unify at $Q=M_{\rm GUT}\simeq 2\times
10^{16}$~GeV, as is expected in the simplest grand unified theories
(GUTs).  On the astrophysical side, SUSY theories with a conserved
$R$-parity offer several candidates for the dark matter particle, among these:
the lightest neutralino, the gravitino and the axion/axino
supermultiplet.

Along with these successes, {\it generic} SUSY models also lead to new
puzzles not present in the Standard Model (SM). Additional (usually
discrete) symmetries are necessary to prevent a too rapid rate for
proton decay. 
Moreover, unless weak scale soft SUSY breaking (SSB) parameters of
the MSSM are flavor-blind \cite{dg} or ``aligned'' \cite{seiberg}, and
nearly real, the model leads to unacceptably large rates for
flavor-changing transitions and $CP$ violating effects.  We stress that
various (sometimes, admittedly {\it ad hoc}) mechanisms have been suggested to
ameliorate these undesired effects which, we speculate, arise because of
our lack of understanding of how SM superpartners feel the effects of
SUSY breaking.

Another possibility to suppress unwanted flavor-changing and
$CP$-violating effects is to arrange for matter sparticles to be heavy
so that their effects are sufficiently suppressed \cite{dine}. Large
matter scalar masses also serve to suppress proton decay processes in a
SUSY GUT.  This {\it decoupling solution} to the SUSY flavor problem usually
requires SSB terms of order 10--100~TeV \cite{arkmur}, 
well beyond expectations from
naturalness, which favors weak scale soft terms.  It was noted as far back
as 1986 \cite{manuel}, and later again in Ref.~\cite{dgpt}, 
that the stability of
the weak scale to radiative corrections requires only the
electroweak (EW) gauginos and third generation sparticles --- these couple with
large strength to the Higgs sector --- to have masses up to ${\cal O}(1)$~TeV, 
while gluinos and superpartners of the first two generations
(whose direct couplings to the Higgs sector are very small, so that
these enter naturalness considerations only at the two-loop level) could
well have multi-TeV masses. Since the most stringent
constraints on flavour- and $CP$-violation come from the first two
generations of quarks and leptons, such a mass spectrum potentially
alleviates the SUSY flavor and $CP$ problems without the need for undue
fine-tuning of parameters. This was subsequently developed into the
framework referred to as ``Effective Supersymmetry'' (ESUSY),
by Cohen, Kaplan and Nelson~\cite{ckn}, who suggested two different
realizations of the split matter generations idea by introducing a
SUSY-breaking sector to which the first two generations couple more
strongly than the third generation.  As a result, the matter scalars of
the first two generations acquire larger SUSY-breaking masses than third
generation scalars.

The ESUSY scenario seems in recent years to have become less favored due
to two measurements. The first --- the measured (2--3)$\sigma$ deviation
in $(g-2)_\mu$ from its Standard Model (SM) expectation~\cite{gm2} ---
seems to require smuons/muon sneutrinos in the sub-TeV range if the
deviation is to be attributed to SUSY. The second --- the increasingly
precise measurement of the dark matter (DM) density of the Universe ---
is difficult to reconcile in ESUSY if the dark matter is assumed to be
dominantly composed of thermal relic neutralinos left over in standard
Big Bang cosmology. With scalars in the multi-TeV range, along with a
bino-like neutralino, the relic density is calculated to be typically
several orders of magnitude higher than the experimentally observed
value~\cite{Jarosik:2010iu}, \be \Omega_{\rm DM} h^2 = 0.1123\pm 0.0035,
\ee although there are some special parameter regions where this need
not be the case.

At the present time, it is not completely clear whether the $(g-2)_\mu$
anomaly is real.  The current discrepancy arises if one adopts the (more
direct to use) hadronic vacuum polarization amplitude from low energy
$e^+e^-\to hadrons$ data. If instead the vacuum polarization is taken
from $\tau$ lepton decay data, then the discrepancy is smaller (but
recently growing!).

In the case of neutralino DM, it is possible to have a small
superpotential $\mu$ term co-existing with large scale masses, as occurs
{\it e.g.} in the hyperbolic branch/focus point region of the
mSUGRA\footnote{Minimal Supersymmetric model with Universal soft terms
at the Gut scale and RAdiative electroweak symmetry breaking.}
model.  Then the lightest neutralino, instead of being typically
bino-like with a small annihilation cross section and concomitantly
large relic density, becomes mixed higgsino dark matter with thermal
relic neutralinos from the Big Bang making up the 
observed cold dark matter relic density.
Other possibilities for obtaining the right relic density
are resonant annihilation through the pseudoscalar Higgs, or
co-annihilation with third-generation sfermions.

Regions of parameter space where the neutralino relic density is too
large cannot be unequivocally excluded in extensions of the model.  For
instance, if one invokes the Peccei-Quinn-Weinberg-Wilczek (PQWW)
solution to the strong CP problem \cite{PQWW}, then one expects the
presence of an axion/axino supermultiplet in SUSY theories. If the axino
$\ta$ is the lightest SUSY particle (LSP), then $\tz_1\to\ta\gamma$ and
other decay modes are allowed, which can greatly reduce the DM abundance
far below the level expected from neutralinos. The scenario of mixed
axion/axino DM has been examined in the SUGRA context recently in
Ref.~\cite{bbs} in a general 19 parameter MSSM, and appears
to be at least as viable as the case with neutralino DM.

In light of these considerations, we feel it would be fruitful to
re-visit some of the phenomenological implications of ESUSY models at
the beginning of the LHC era. In our analysis, we subsume the
qualitative features of the ESUSY model: {\it i.e.} third generation,
Higgs sector and EW gaugino masses at or below the TeV scale, with
multi-TeV SSB parameters for the first two generations. For simplicity
we also assume gaugino mass unification.\footnote{While the gluino mass
parameter can, in principle, be hierarchically larger, RGE effects due
to a very heavy gluino would raise third generation squark mass
parameters to high values. Gauge coupling unification also prefers that
the gluino not be too heavy.}  While we adopt the qualitative picture of
Effective SUSY, we do {\it not} assume the validity of the more
speculative mechanisms that the authors of Ref.~\cite{ckn} suggest for
the hierarchy between the SSB parameters of the first two generations
and the corresponding parameters for the third generation and the
gaugino sector.  For our phenomenological analysis, we simply assume
that such a hierarchy occurs for SSB parameters renormalized at a high
scale that we take to be $M_{\rm GUT}$. We also take the less ambitious
view that the SUSY flavour and $CP$ problems are ameliorated, but not
completely solved, by this hierarchy, and allow the first two
generations of scalars to have masses in the 5--20~TeV range, so that
some degree of universality/alignment (for the first two generations) is
still necessary to satisfy the most stringent flavour constraints. For
simplicity, we will assume GUT scale scalar mass universality in the
subspace of the first two generations, but allow independent SSB
parameters for the third generation and the Higgs sector. This
assumption of diagonal GUT scale scalar SSB mass squared matrices
undoubtedly has an effect on flavor physics~\cite{bfac}, but should have
very limited impact on the implications for collider physics and
cosmology of the ESUSY scenario --- which are after all our main interest
in this paper.

The ESUSY model parameter space we will examine is thus given by
the set of parameters (renormalized at the GUT scale) 
\be
m_0(1,2),\ m_0(3), m_{H_u},\ m_{H_d},\ A_0,\ m_{1/2},\ \tan\beta ,\ {\rm
sign}(\mu )
\label{eq:pspace}
\ee 
along with the top quark mass $m_t=173.1\pm 1.3$~GeV. 
Here $m_0(1,2)$ and $m_0(3)$ are the masses of the
first/second and of the third generation sfermions, respectively;
$m_{H_u}\equiv m_{H_u}^2/\sqrt{|m_{H_u}^2|}$ and  
$m_{H_d}\equiv m_{H_d}^2/\sqrt{|m_{H_d}^2|}$, 
are the SSB mass parameters of the up- and down-type
Higgs scalars; $A_0$ is a universal trilinear coupling (relevant mostly only
for the third generation); $m_{1/2}$ a universal
gaugino mass parameter; and $\tan\beta\equiv v_u/v_d$. Radiative electroweak
symmetry breaking can be used to determine $\mu^2$ using the measured
value of $M_Z$. 

We wish to maintain the successful gauge coupling unification at
$Q=M_{\rm GUT}$, so that we will assume here that the MSSM is the
correct effective field theory between $M_{\rm GUT}$ and
$\widetilde{M}=m_0(1,2)$, the scale at which first and second generation
scalars decouple from the theory. Below $\widetilde{M}$, we have ESUSY
--- the
SM with two Higgs boson doublets together with third generation scalars,
gauginos and higgsinos --- as the effective theory.

Models with third generation scalar and gaugino-higgsino
sector at $\alt$TeV but multi-TeV first and second generation mass 
parameters have been investigated in the past. 
Shortly after Cohen {\it et al.}~\cite{ckn} 
laid down their framework, it was noted
that two loop RGE effects arising due to heavy first/second generation
scalars act to suppress third generation scalar mass parameters (even
to tachyonic values), and a variety of flavor and $CP$-violating
constraints were examined~\cite{arkmur,hisano}.  
In Refs.~\cite{feng,imh} it was shown that the inverted scalar
mass hierarchy which is the hallmark of  the
ESUSY scenario could emerge dynamically in models with Yukawa
coupling unification, although the  $\widetilde{M}:M_{\rm weak}$ ratio 
was found to be limited~\cite{imh}. This class of models --- requiring
$t-b-\tau$ Yukawa coupling unification and $SO(10)$ like boundary
conditions with  $\tan\beta\sim 50$ --- has been
investigated in detail in Refs.~\cite{yuk,bdr}.  
In Ref.~\cite{gsimh}, the magnitude of the $\widetilde{M}:M_{\rm weak}$ 
hierarchy was investigated in models without Yukawa coupling unification
where the matter scalar mass
parameters are already taken to be split at the GUT scale.  This study
assumed equal GUT scale values of third generation and Higgs SSB masses,
and as a result, was limited in scope compared to the results to be
presented here.  In addition, several authors have investigated other
related aspects of supersymmetric models with heavy
scalars~\cite{biswarup}. 

The remainder of this paper is organized as follows. In
Sec.~\ref{sec:pspace}, we explore the parameter space in
Eq.~(\ref{eq:pspace}) and map out ranges of parameters that potentially
lead to ESUSY at the weak scale. In Sec.~\ref{sec:mcmc}, we perform a
Markov Chain Monte Carlo (MCMC) analysis to search for SUSY scenarios
which fulfill the ESUSY conditions, subject to various experimental
constraints.  We first describe the setup of our MCMC, and then show
results for posterior probability distributions both for input
parameters that result in ESUSY, as well as for expectations for various
sparticle masses and selected experimental observables. 
Moreover, we pick out some characteristic benchmark points for further study. 
In Sec.~\ref{sec:lhc}, we first present the main particle production rates
and decay patterns relevant for LHC phenomenology of the ESUSY
scenarios, and then illustrate the diversity of LHC phenomena through
more detailed discussion of the phenomenology of the benchmark points,
including the possibility that the lightest MSSM sparticle (LMP, as distinct
from LSP) may be charged or coloured. 
We conclude in Sec. \ref{sec:conclude} with a summary of our results.

\section{Viable spectra and parameter space of the ESUSY model}
\label{sec:pspace}


Constrained supersymmetric models such as the mSUGRA model have, for a
fixed gluino mass, an upper limit on how massive scalars can be. For a
model with universal scalar mass soft terms at the GUT scale, given by
$m_0$, with $m_{1/2}$ fixed, the weak scale value of
$\mu^2$ diminishes as $m_0$ increases.  Ultimately, when $m_0$ is large enough, $|\mu |$
becomes comparable to the weak scale value of the bino mass $M_1$, and
one gains neutralinos of mixed higgsino-bino composition which make a
good candidate for thermal WIMPs.  This is the so-called hyperbolic
branch/focus point (HB/FP) region of the SUGRA parameter space~\cite{hb_fp}. 
As $m_0$
increases even further, $|\mu |$ decreases even more and the $\tz_1$
becomes nearly pure higgsino, and the higgsino-like lightest chargino
may drop below limits from LEP2 searches. For yet higher values of
$m_0$, $\mu^2$ becomes negative, signaling an inappropriate
electroweak symmetry breaking (EWSB) pattern.  Thus, in the mSUGRA model, for a
given value of $m_{1/2}$, $m_0$ can become no larger than a few TeV.

\begin{figure}[t]
\begin{center}
\epsfig{file=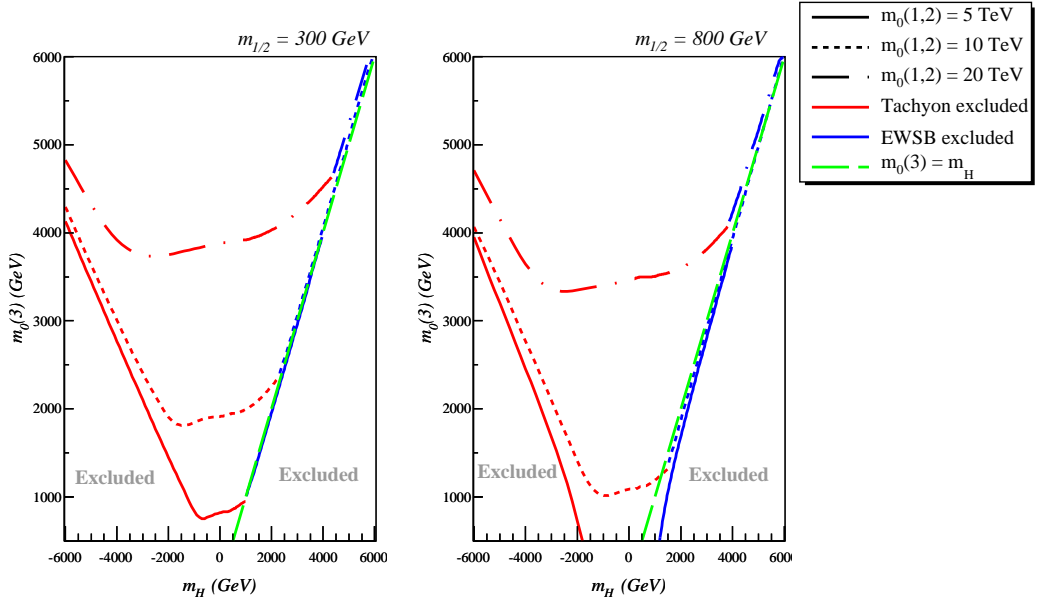,width=14cm}
\end{center}
\vspace*{-4mm}
\caption{\it Regions of parameter space leading to viable sparticle mass
spectra in the $m_H\ vs.\ m_0(3)$ plane for $m_0(1,2)=5,\ 10$ and 20
TeV. We fix $A_0=0$, $\tan\beta=10$ and $\mu>0$ and choose $m_{1/2}=300$
GeV in the left frame and $m_{1/2}=800$ GeV in the right frame.  The
region below the red lines is excluded because of tachyonic third
generation masses or, for large negative values of $m_H$ because
$m_A^2<0$, while the region to the right of the blue lines does not
allow EWSB. The green dashed line has $m_0(3) = m_H$.  The ESUSY spectra
results from taking $m_0(3)$ to its lowest allowed values.
}\label{fig:mH}
\end{figure}

If we proceed to models with non-universal generations, taking
$m_0(1,2)$ as independent of $m_0(3)$, with
$m_0(3)=m_{H_{u}}=m_{H_{d}}\equiv m_H$ as in Ref.~\cite{gsimh}, we
expect, using 1-loop RGEs, by the same reasoning, an upper bound on
$m_0(3)$.  The point is that the first and second generations
essentially decouple from the RG evolution of the Higgs SSB
parameters (which enter the electroweak potential minimization
conditions, see Ref.~\cite{wss}).  At two loop order, however, very large
scalar masses do affect the running of the other scalar masses and in
fact lead to their suppression at the weak scale~\cite{hisano,gsimh}.  
As $m_0(3)$ decreases (or $m_0(1,2)$ increases), so $m_0(3) \ll m_0(1,2)$,
one of the third generation scalars becomes the LSP, turning tachyonic
for even smaller (larger) $m_0(3)$ ($m_0(1,2)$) values. Therefore, with
$m_0^2(3)=m_{H_{u,d}}^2$ at the GUT scale, there are both an upper (from
$\mu^2 >0$) and a lower (from $m_{\tst,\tb,\tilde{\tau}}^2 > 0$) bound on
$m_0(3)$.  In Fig.~\ref{fig:mH}, we show regions of allowed parameter
space in the $m_H$ versus $m_0(3)$ plane for $m_0(1,2)=5$, 10 and 20~TeV,
with {\it a}) $m_{1/2}=300$ GeV and {\it b}) $m_{1/2}=800$ GeV, where
the spectrum calculations are performed using {\tt
  Isajet7.80}~\cite{isa78}.
The green 
dashed line is where $m_H\equiv m_0(3)$.  The region to the right of the blue
lines is forbidden due to lack of EWSB, while the region below the red
lines is forbidden because a third generation scalar becomes tachyonic.
The region to the left and above the contours yields viable spectra.  

For low $m_{1/2}$, we see from Fig.~\ref{fig:mH}{\it a}) that if we
restrict our attention to $m_H=m_0(3)$ (green dashed line) only the
interval 1 TeV $\lesssim m_0(3) = m_{H} \lesssim 4$ TeV is allowed for
$m_0(1,2) = 5$ TeV, while there are no allowed values for $m_0(1,2) =
10$ or 20 TeV, since for these cases the $m_0(3) = m_{H}$ line lies
entirely in the excluded region. However, through 1-loop RGE running
effects, an increase in $m_{1/2}$ increases $m_{\tq}$. Hence larger
values of $m_0(1,2)$ become viable if $m_{1/2}$ is large enough. This is
seen in Fig.~\ref{fig:mH}{\it b}), where the green dashed line (again
representing $m_0(3) = m_{H}$) shows that, for $m_{1/2} = 800$ GeV, the
range of allowed $m_0(3) = m_{H}$ values is larger for $m_0(1,2) = 5$
TeV, where the tachyonic lower bound is gone, so $0 < m_0(3) = m_{H}
\lesssim 5$ TeV. Besides, if a neutralino LSP is required, we have $0.25
< m_0(3) = m_{H} \lesssim 5$~TeV.  For the higher $m_0(1,2)$ cases we
see that the increase in $m_{1/2}$ provides an allowed region for
$m_0(1,2) = 10$ TeV, but still is insufficient for the $m_0(1,2) = 20$
TeV case. It is for this reason that the solutions found in
Ref.~\cite{gsimh} with very large $m_0(1,2)$ values, as needed to
suppress FCNCs, also required rather high $m_{1/2}$ values, and
consequently very heavy gluinos: quite beyond the reach of LHC.

Since there is no compelling theoretical argument to link the Higgs and
third generation mass scales, it seems reasonable 
to adopt independent values of $m_0(3)$ and $m_{H_{u}}=m_{H_d}= m_H$
(NUHM1) or $m_{{H_u}}\ne m_{H_d}$ (NUHM2). In this case, for a large
value of $m_0(3)$, we can take $m_H^2\ll m_0^2(3)$, and thus give $m_H$
a head start on running towards the necessary negative squared values
which are needed for successful EWSB (recall that at the weak scale,
$\mu^2\simeq -m_{H_u}^2$ for even modest values of $\tan\beta$). This
is the well-known feature of NUHM1 models wherein increasing $m_H$
results in a decrease of the weak scale value of $|\mu
|$~\cite{nuhm}. Note that  large negative values of $m_H$ are
excluded because there $m_A^2<0$, signaling that EWSB is
not correctly obtained. 

If we adopt a low value of $|m_H|\alt 1$ TeV, and then move from high to
low $m_0(3)$ values, the third generation scalars decrease in mass.  The
region for ESUSY with third generation scalars at $\alt$TeV is at
$m_0(3)$ values just above the tachyonic third generation region. Thus,
the strategy for gaining viable ESUSY spectra is to 1.)~adopt a large
value of $m_0(1,2)\sim 5-20$ TeV, then 2.)~adopt a low value of $m_H\sim
0-2$ TeV, and finally, 3.)~starting at a several TeV value of $m_0(3)$,
decrease its value until third generation scalars dip below the TeV
region. If we want low $|\mu |$ values as well, then increase $m_H$ as
close to the EWSB boundary (on the right of the figure) as desired.

\begin{figure}[t]
\begin{center}
\epsfig{file=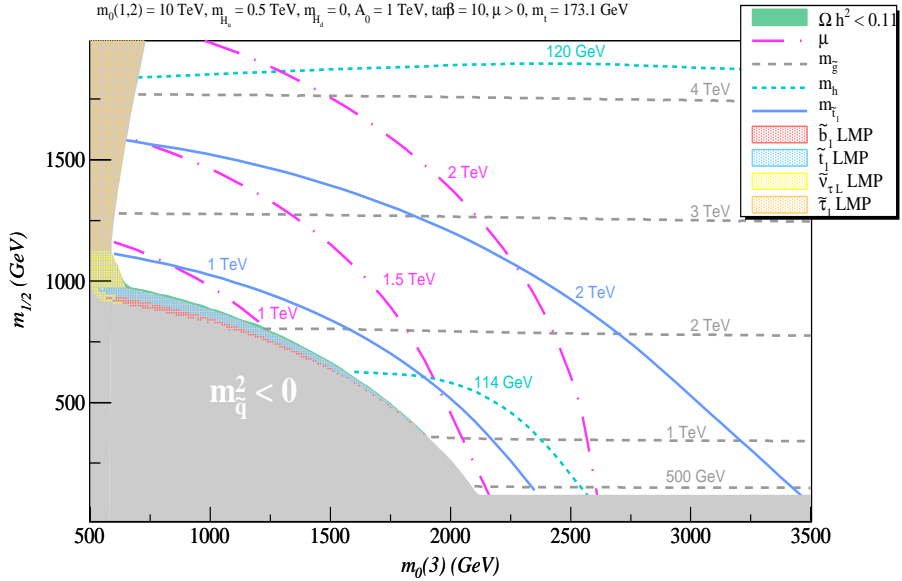,width=12cm,height=8cm}
\end{center}
\vspace*{-4mm}
\caption{\it Contours of $\mu$, gluino, stop and Higgs ($h$) boson
masses in the $m_0(3)\ vs.\ m_{1/2}$ plane for $m_0(1,2)=10$ TeV,
$m_{H_{u}}= 0.5$ TeV, $m_{H_{d}}=0$, $A_0=1$ TeV, $\tan\beta =10$ and
$\mu >0$. Also shown are regions where various scalars are the lightest 
MSSM particles (LMPs). 
In the almost impossible to see narrow green band whose thickness
is $\alt 10$~GeV that hugs the upper boundary of the stop LMP region,
the thermal relic abundance $\Omega_{\tz_1}h^2<0.11$.  The gray shaded
region is excluded because either we have tachyonic sparticles or the
chargino mass is below the corresponding LEP2 limit.  }\label{fig:plane}
\end{figure}

It is instructive to portray ESUSY in various parameter
planes. Figure~\ref{fig:plane} shows the allowed region in the $m_0(3)\
vs.\ m_{1/2}$ plane with $m_0(1,2)=10$ TeV, $m_{H_u}= 0.5$ TeV,
$m_{H_d}= 0$ TeV (along with $A_0=1$ TeV, $\tan\beta =10$ and $\mu >0$).
The left-side gray shaded region gives tachyonic stop or sbottom masses,
while in the lower-right gray region the chargino mass is below its LEP2
limit.  The white region gives viable supersymmetric mass spectra. We
plot contours of $m_{\tg}$ (gray dashed), $m_{\tst_1}$ (blue solid),
$m_h$ (dashed cyan) and $\mu$ (magenta dot-dashed).  The region to the
lower left of the cyan dashed contour has $m_h<114$ GeV. The slim region
adjacent to but right of the tachyonic stop region gives viable ESUSY
spectra with top squark masses $m_{\tst_1}<1$ TeV. There is an
almost-impossible-to-see green region of thickness $\alt 6$~GeV in
$m_{1/2}$ hugging the boundary of the tachyonic stop region where the
thermal neutralino abundance is in accord with measurement:
$\Omega_{\tz_1}h^2<0.11$. This is, in fact, the top squark
co-annihilation region~\cite{stop} for ESUSY. Along the tachyonic boundary is
another narrow region where the $\tst_1$ can become the lightest MSSM
particle (LMP), though not necessarily the LSP if a non-MSSM $R$-odd
sparticle (such as the axino) is lighter.  We also show the regions where
$\tb_1$, $\tilde{\tau}_1$ or $\tilde{\nu}_{\tau L}$ is the
LMP.\footnote{With a common value of $m_0(3)$, the $\tnu_{\tau}$ LMP may
be surprising since at the one-loop level Yukawa interactions drive
$m_{\ttau_R}^2$ to smaller values than $m_{\ttau_L}^2$. In this case,
however, two-loop effects due to heavy scalars are large, and for
smaller values of $m_{1/2}$ can lead to $m_{\ttau_L}< m_{\ttau_R}$ and a
sneutrino LMP because of the $D$-term. Since one-loop effects increase
with $m_{1/2}$, ultimately the usual situation with stau LMP is
obtained. Incidently, the compensation between the one and two-loop
contributions implies that for certain parameters we can have a near
equality of the left- and right-stau mass parameters, and so large stau
mixing even though the tau Yukawa coupling is not particularly large.}

\begin{figure}[ht!]
\begin{center}
\epsfig{file=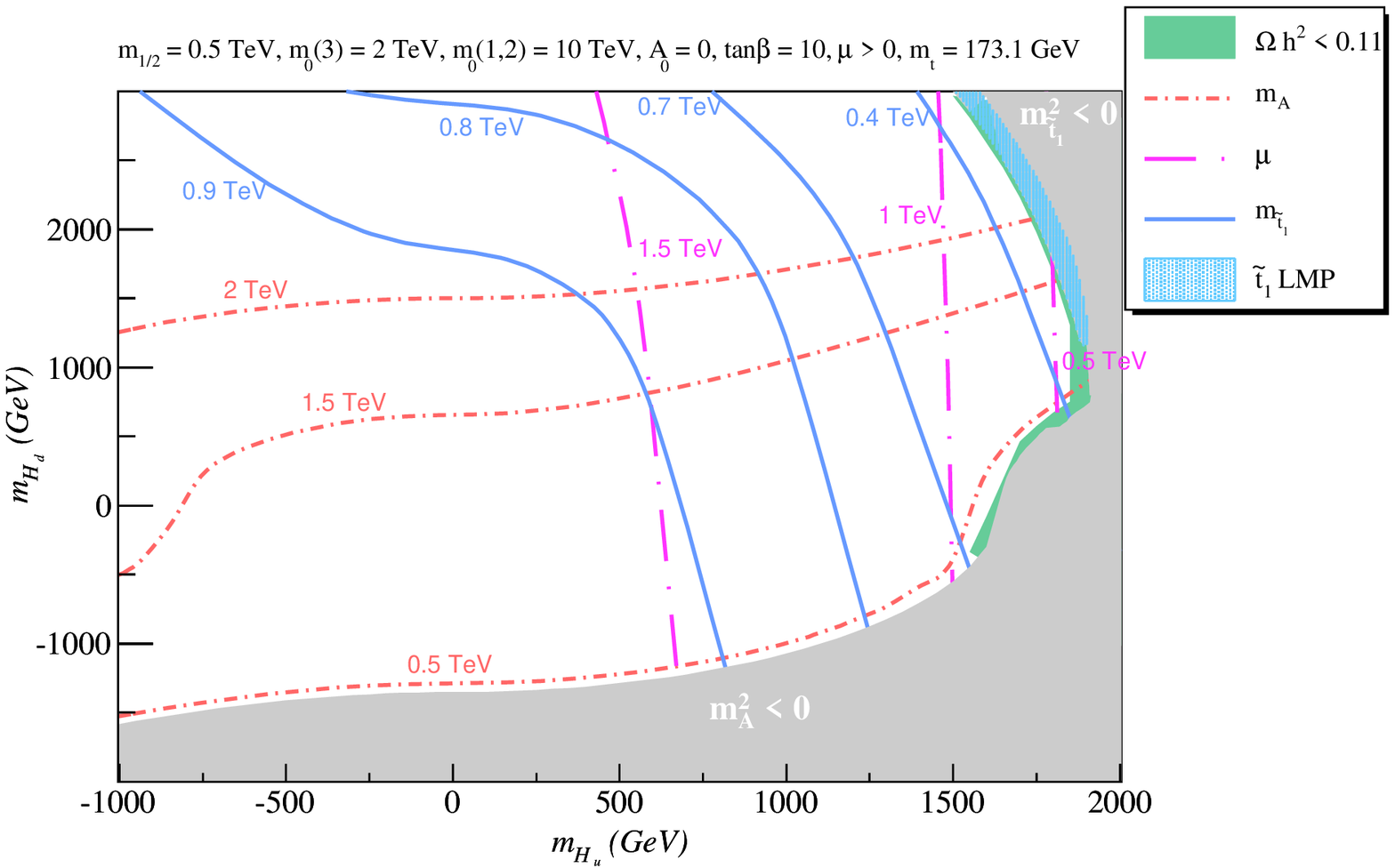,width=11cm,height=7cm} \vspace{2pt}
\epsfig{file=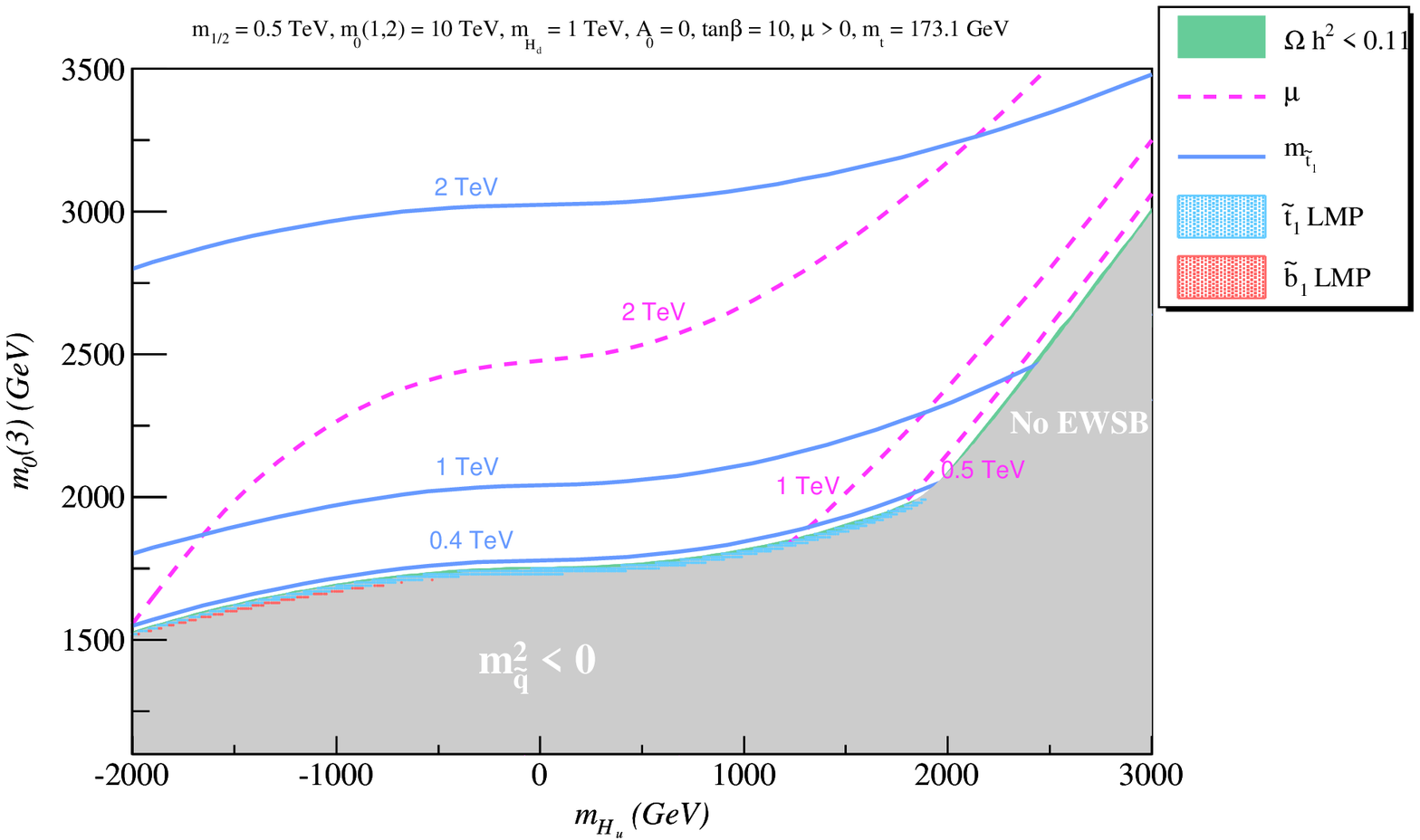,width=11cm,height=7cm}
\end{center}
\vspace*{-4mm}
\caption{\it The uppper frame a) shows contours of $\mu$,
$m_{\tilde{t}_1}$ and $m_A$ in the $m_{H_u} \ vs.\ m_{H_d}$ plane for
$m_0(1,2)=10$ TeV, $m_0(3) = 3$ TeV, $m_{1/2}= 0.5$ TeV, $A_0=0$,
$\tan\beta =10$ and $\mu >0$. In the green band, the thermal relic abundance
$\Omega_{\tz_1}h^2<0.11$ and in the shaded blue region the $\tst_1$ is
the lightest MSSM particle.  The gray shaded region is excluded as
explained in the text.  The lower frame b), shows contours of $\mu$ and
$m_{\tilde{t}_1}$ in the $m_{H_u}\ vs.\ m_0(3)$ plane for $m_0(1,2)=10$
TeV, $m_{1/2}= 0.5$ TeV, $m_{H_d} = 1$ TeV, $A_0=0$, $\tan\beta =10$ and
$\mu >0$.  }\label{fig:planeB}
\end{figure}

In Fig.~\ref{fig:planeB}{\it a}), we fix instead the values of
$m_0(1,2)$, $m_0(3)$ and $m_{1/2}$ and plot contours of $m_{\tst_1}$,
$\mu$ and $m_A$ in the $m_{H_u}\ vs.\ m_{H_d}$ plane.  As discussed
above, for intermediate positive ranges of $m_{H_d}$ in the plot, where
$m_{H_u} \sim m_{H_d}$, as $m_{H_u}$ increases, the value of $\mu$
decreases until $\mu^2 < 0$ and EWSB is no longer realized. This can be
seen by the $\mu$ contours in this part of the figure where we also see
a significant green region where the thermal neutralino relic density is
consistent with its observed value.  In the lower portion (low
$m_{H_d}$) of the grey area on the right and at the bottom in
Fig.~\ref{fig:planeB}{\it a}), we do not obtain radiative EWSB because
$m_h^2$ or $m_A^2$ become negative. The green region in this vicinity is
where the neutralino relic density is compatible because of higgs
resonance annihilation~\cite{Afunnel}. Finally, for large enough
positive values
of $m_{H_u}$ and $m_{H_d}$, $\tilde{t}_1$ becomes the LMP (shaded blue
region), with $m_{\tilde{t}}^2 < 0$ above that (upper portion of the
grey area on the right) and co-annihilation with stops leads to a compatible
neutralino relic density.  Finally, we present in
Fig.~\ref{fig:planeB}{\it b}) the allowed region in the $m_{H_u}\ vs.\
m_0(3)$ plane with $m_0(1,2)=10$ TeV, $m_{1/2} = 0.5$ TeV and $m_{H_d}=
1$ TeV. We see that the ESUSY scenario is realized at low $m_0(3)$ and
$m_{H_u} \lesssim m_0(3)$.

\section{Markov Chain Monte Carlo analysis of ESUSY}
\label{sec:mcmc}

In this section, we perform an MCMC analysis to map out the regions 
of ESUSY parameter space that are consistent with current experimental 
constraints from collider and flavour physics. The characteristics  
of ESUSY are enforced by means of theoretical priors.   
We first briefly describe our set up 
and list the likelihood functions and model priors that we use, and
then show our results for the posterior probability distributions both
for input parameters as well as various sparticle masses and selected
experimental observables. In Sec.~\ref{ssec:BM}, we present some
benchmark points for more detailed study.

\subsection{Setup of the MCMC scans} \label{ssec:scans}

The setup and procedure of our MCMC closely follow~\cite{Belanger:2009ti}, 
so we do not repeat these here but instead just give the key data of the analysis. 
For details on the MCMC method and Bayesian analysis, see 
{\it e.g.}~\cite{de Austri:2006pe,Allanach:2007qk,Trotta:2008bp}. 
In addition to {\tt Isajet7.80}~\cite{isa78}, which we use for the spectrum
calculation, we use {\tt SuperIso2.7}~\cite{Mahmoudi:2008tp} for
computing B-physics observables and {\tt
micrOMEGAs2.4}~\cite{Belanger:2006is} for the calculation of the
neutralino LMP abundance.

For each point $P$ in the scan of ESUSY parameter space, we first compute individual likelihoods $L(O_i)$ for each experimental observable $O_i$.  Then the overall likelihood is given as the product of these individual likelihoods as $L_P=\prod L(O_i)$.
The observables we use are listed in Table~\ref{tab:observables} along with the parameters of the corresponding likelihood
functions used. The forms of these likelihood functions are given by,
\begin{equation}
  L_1(x,x_0,\sigma_x) = \frac{1}{1+\exp[-(x-x_0)/\sigma_x]}, \quad 
  L_2(x,x_0,\sigma_x) = \exp\left[-\,\frac{(x-x_0)^2}{2\,\sigma_x^2}\,\right] \,,
\end{equation}  
for observables for which there is only an upper or lower bound, 
and for observables for which a measurement is available, respectively.
For the LEP limit on the Higgs mass, we use~\cite{Buchmueller:2009fn}  
\begin{equation}
  \chi^2(m_h) = \frac{(m_h-m_h^{\rm limit})^2}{(1.1~{\rm GeV})^2+(1.5~{\rm GeV})^2}
\end{equation}
with $m_h^{\rm limit}=115$~GeV. The likelihood $L(m_h)$ is then given by 
$L(m_h)=e^{-\chi^2(m_h)/2}$ for $m_h<115$~ GeV and $L(m_h)=1$ 
for $m_h\ge 115$~GeV.

\begin{table}[t]
\begin{center}
\begin{tabular}{||l|c|l|l||}\hline\hline
Observable & Limit & Likelihood function & Ref. \\ 
\hline
  ${\rm BF}(b\to s\gamma)$ & $(3.52 \pm 0.34) \times 10^{-4}$ & 
  $L_2(x,3.52 \times 10^{-4}, 0.34  \times 10^{-4})$& \cite{Barberio:2008fa,Misiak:2006zs} \\ 
\hline
  ${\rm BF}(B_s\to \mu^+\mu^-)$ & $\le 5.8 \times 10^{-8}$ & 
  $L_1(x,5.8 \times10^{-8}, -5.8 \times 10^{-10})$ & \cite{Aaltonen:2007kv} \\ 
\hline
  $R(B_u\to \tau\nu_\tau)$ & $1.28\pm0.38$ & $L_2(x,1.28,0.38)$ & \cite{Barberio:2008fa} \\
\hline
  $m_t$ & $173.1\pm 1.3$ & $L_2(x,173.1,1.3)$ & \cite{tev:2009ec}\\
\hline
  $m_h$ & $\ge 114.5$ &  $1$ or $\exp(-\chi^2(m_h)/2)$ & \cite{Buchmueller:2009fn}  \\ 
\hline 
SUSY mass limits & LEP limits & $1$ or $10^{-9}$ & \cite{lepsusy} \\
\hline
\hline
\end{tabular}
\end{center}
\caption{\label{tab:observables} Observables used in the likelihood calculation.}
\end{table}

In order to favour (sub)TeV-scale electroweak gauginos, higgsinos, and third generation 
sfermions, we multiply the likelihood obtained from the experimental constraints  
with the following model prior:
\begin{equation}
   L_{M_{\rm eff}}(m_X)= \frac{1}{1+\exp((m_{X}-M_{\rm eff})/{170 \ {\rm GeV}} ) }  
   \label{eq:m3prior}
\end{equation}
for each $X=\tilde\chi^+_1,\,\tilde\chi^+_2,\,\tilde t_1,\,\tilde t_2,\, \tilde b_1$.
Choosing, for instance, $M_{\rm eff}=1.5$~TeV, we get $L_{M_{\rm eff}}=0.95$, $0.5$, 
$0.05$ for $m_X=1$, $1.5$ and $2$~TeV respectively; 
for $M_{\rm eff}=1$~TeV, we get $L_{M_{\rm eff}}=0.95$, $0.5$, $0.05$ for 
$m_X=0.5$, $1$, $1.5$~TeV respectively. We have run chains for different 
$M_{\rm eff}$ and checked that the effect is indeed only to vary the upper 
bound on the effective SUSY masses; the qualitative features of the parameter 
space remain unchanged. 

Finally, regarding the model parameters, we allow $m_{1/2}=[0,\, 2]$~TeV, 
$m_{0}(3) = [0,\, 10]$ TeV, 
$m_{H_{u,d}}^{} =[-10,\, 10]$~TeV (with $m_{H_{u,d}}$ unrelated), 
$A_{0} = [-40,\, 40]$~TeV, and 
$\tan\beta = [2,\, 60]$, taking flat prior probability distributions for all 
these parameters. 
In addition we let the top mass vary within 
$m_t=173.1 \pm 1.3$ GeV~\cite{Group:2009qk} with a Gaussian distribution. 
For $m_0(1,2)$, we show results for two different approaches:

\begin{enumerate}
\item we let $m_0(1,2)$ vary from 5 to 20 TeV with a 
flat prior probability distribution ($L_{\widetilde M}\equiv 1$), 
or
\item we let $m_0(1,2)$ vary without limits but apply a model prior of  
\begin{equation}
   L_{\widetilde M}(m_0(1,2))= \frac{1}{1+\exp((10-m_0(1,2))/{1.7 \
   {\rm TeV}})} \, , 
   \label{eq:m0prior}
\end{equation}
in order to favour $m_0(1,2)$ values beyond 10 TeV.
\end{enumerate}
The total likelihood of a point  is then taken to be
$L_{\rm tot}=L_P\times L_{M_{\rm eff}}\times L_{\widetilde M}$.

Before turning to the results, a comment is in order regarding the
nature of the LMP.  If we do not impose any dark matter requirement on
the LMP, the large majority of points have a neutralino LMP with a far
too high relic density of up to $\Omega_\chi h^2\sim 10^3$. As noted in
the Introduction, this can be in agreement with cosmological
observations if the ``true'' LSP is actually an axino, {\it i.e.}\ in
the case of mixed axion/axino dark matter~\cite{bbs,Baer:2008eq}.  
Moreover, about 12\% (4\%) of the points accepted by the chains with $M_{\rm
eff}=1$~($1.5$)~TeV have a stop, sbottom or stau LMP. Again these points
may be viable if the true LSP is the axino. However, as we will discuss
below, the phenomenology in these charged LMP cases is quite different
from the neutralino LMP case: in particular, if the stop, sbottom or
stau LMP does not decay promptly but gives a charged track. In what
follows, we will confine our MCMC analysis to the case of a neutralino
LMP,  which indeed occurs most frequently.
We will, however, comment on the phenomenology of the colored or 
charged LMP case at the end of the next section.

\subsection{MCMC results with neutralino LMP}
\label{subsec:res}

\begin{figure}[t]
\begin{center}
\epsfig{file=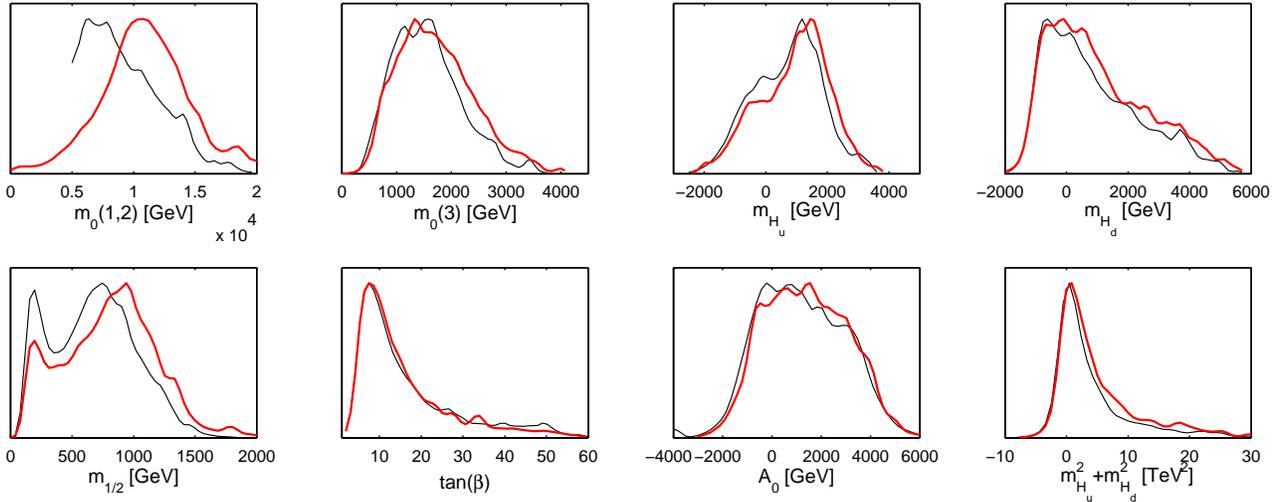,width=17cm}
\end{center}
\vspace*{-4mm}
\caption{\it Posterior probability distributions of the input parameters
of the ESUSY model from MCMC scans with $M_{\rm eff}=1$~TeV, comparing
the two priors for $m_0(1,2)$. We take $\mu>0$, and assume that the
neutralino is the LMP. The thin black lines are
for case 1.\ where $m_0(1,2)=5-20$~TeV with a flat distribution, while
the thick red lines are for case 2.\ which has no limit on $m_0(1,2)$
but the prior Eq.~(\ref{eq:m0prior}) disfavours low values for this
parameter. The last panel shows the combination $m_{H_u}^2+m_{H_d}^2$.
\label{fig:mcmc1}}
\end{figure}

Figure~\ref{fig:mcmc1} shows the posterior probability distributions of
the input parameters
of the
ESUSY model from MCMC scans requiring a neutralino LMP.  Here we have
used $M_{\rm eff}=1$~TeV in Eq.~(\ref{eq:m3prior}).  Unseen dimensions
are marginalized over.  The figure compares the two priors for
$m_0(1,2)$: The thin black lines are for case 1. where
$m_0(1,2)=5-20$~TeV with a flat distribution, while the thick red lines
are for case 2. which has no limits on $m_0(1,2)$ but the choice of the
prior in Eq.~(\ref{eq:m0prior}) favours higher values of $m_0(1,2)$. We
see that, as anticipated, we find solutions where third generation
sfermion and gaugino SSB parameters are typically 1--2~TeV, while the
first/second generation SSB parameters are favoured to lie beyond
10~TeV. Moderate values of $\tan\beta$ are favoured. 

As can be seen, the precise condition on $m_0(1,2)$ hardly influences
the probability distributions of the other parameters.  A possible
exception is $m_{1/2}$, which features a double-peak distribution.  This
comes from the fact that at small $m_{1/2}$, the parameter space volume
is constrained in the directions of the scalar mass parameters, while
there is more space in the $\tan\beta$ direction. For large $m_{1/2}$,
on the other hand, $\tan\beta$ is very much constrained by BR$(b\to
s\gamma)$, while there is more volume in the large $m_0(1,2)$
directions.  The effect is more pronounced for case 1. which does not
disfavour smaller values for $m_0(1,2)$.  Moreover, we see that larger
$m_0(1,2)$ prefers somewhat larger $m_{1/2}$.\footnote{The shape of the
posterior probability distribution of $m_{1/2}$, and $\tan\beta$ are
likely the most sensitive to the $b\to s\gamma$ constraint, and hence to
our assumption that there are no off-diagonal pieces in the GUT scale
squark mass matrices.}  It is also interesting to note that very high
$m_0(1,2)$ around 15--20 TeV suffers from a low probability. This is
because as $m_0(1,2)$ increases, we are forced to increasing larger
values of $m_0(3)$ (see Fig.~\ref{fig:mH}). For values of $m_0(3)$ in the
multi-TeV range, third generation SSB parameters at $\alt$TeV can occur only
when the large two loop RGE contributions  rather precisely cancel the
naturally multi-TeV contribution that we start with -- too
little cancellations leave large SSB parameters, while too much
cancellation leads to tachyonic masses. As a result, the region
of parameter space with a light third
generation rapidly shrinks when $m_0(1,2)$ exceeds $\sim$ 15--20~TeV. 

\begin{figure}[t]
\begin{center}
\epsfig{file=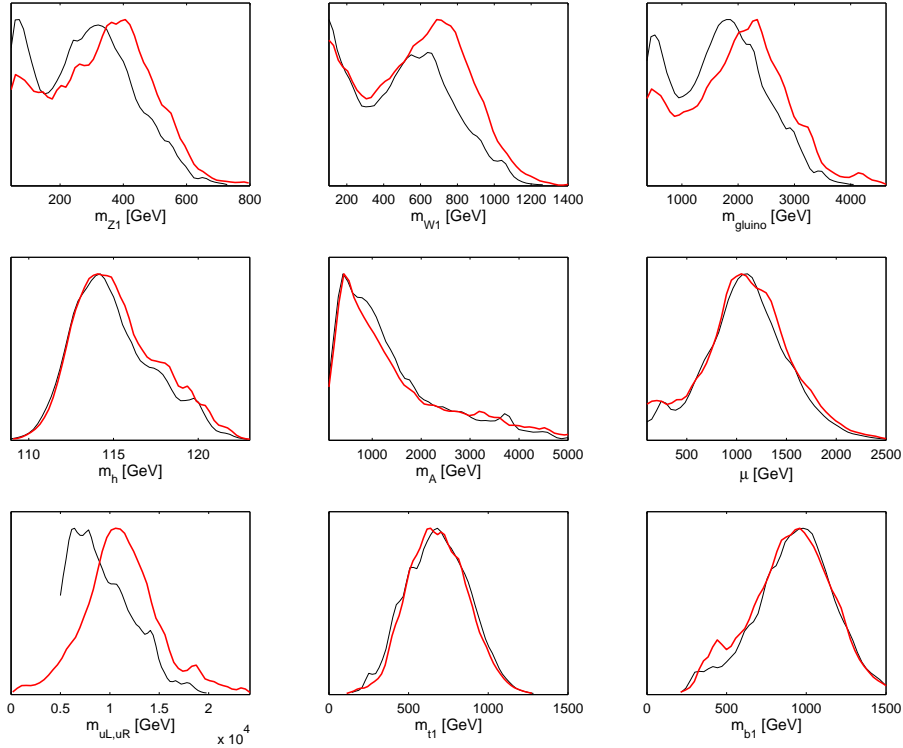,width=12cm}
\end{center}
\vspace*{-4mm}
\caption{\it Same as Fig.~\ref{fig:mcmc1} but for various sparticle
  and Higgs masses, and $\mu$.
}\label{fig:mcmc2} 
\end{figure}

The posterior probability distributions of several relevant masses are shown in
Fig.~\ref{fig:mcmc2}.  We infer from the EW gaugino spectrum 
(and also the $\mu$) distribution that most of the time the lightest
neutralino is bino-like. 

\begin{figure}[t]
\begin{center}
\epsfig{file=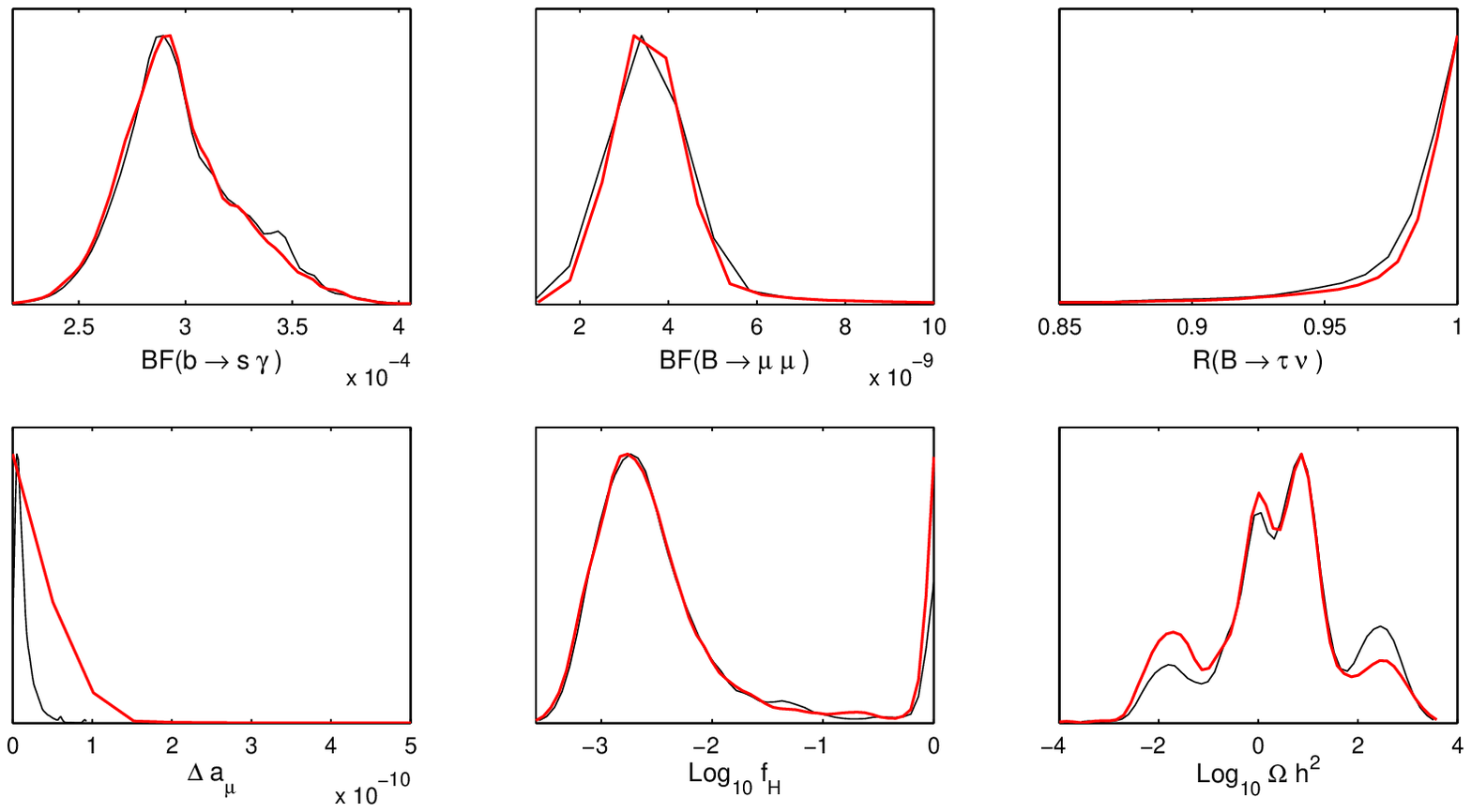,width=12cm}
\end{center}
\vspace*{-4mm}
\caption{\it Same as Fig.~\ref{fig:mcmc1} but for a variety 
of  low energy observables, the higgsino content of the lightest neutralino, 
and the thermal neutralino relic density. 
}\label{fig:mcmc3} 
\end{figure}

Figure~\ref{fig:mcmc3} displays the probability distributions of
$B$-decay observables, $\Delta a_\mu=(g-2)_\mu^{\rm SUSY}$, the higgsino
fraction $f_H = v_1^{(1)2}+v_2^{(1)2}$ (in the notation of Ref.~\cite{wss})
of the $\tilde Z_1$, and the relic abundance $\Omega h^2$ of the $\tilde
Z_1$. It is clear that if the muon anomalous moment is confirmed, it
cannot arise within the ESUSY framework.  We note that there is
non-negligible fraction of the parameter space with a higgsino-like
LMP. We also see that the neutralino relic density peaks around $\Omega
h^2\simeq 1-10$ and goes up to $\Omega h^2\sim 10^4$. Nevertheless,
solutions with low values of $\Omega h^2\alt 0.1$, where the relic
density of thermal neutralinos does not yield a universe that is too
short-lived, also have non-negligible probability. They occur because of
Higgs funnel annihilation, coannihilation with light stops and/or
sbottoms (or, less likely, staus), or because of a large LMP higgsino
component.

\subsection{Effective SUSY benchmark points}
\label{ssec:BM}

To illustrate the types of SUSY spectra and the diverse phenomenology
that can result 
in ESUSY models, we list five benchmark points in Table
\ref{tab:bm}. The first three points we list have a neutralino LMP,
while points ES4 and ES5 have a sbottom and a stau LMP, respectively.
Point ES1 adopts an $m_0(1,2)=10$ TeV, 
which is the most favored value in the probability distributions resulting from the MCMC.  
On the other hand, ES1 has a value of 
$m_{\tg}=524$ GeV, not far from the ESUSY minimum of $m_{\tg}\agt 460$
Gev. The chargino $\tw_1$ and the neutralino $\tz_{2}$ have masses
of just $139$~GeV, and the $\mu$ parameter is 857~GeV.
All  third generation squarks except $\tb_2$, as well as all EW gauginos 
are $\alt$1 TeV scale, while the two staus are over 2 TeV.

In point ES2, $m_0(1,2)$ is increased to 12 TeV.  ES2 has much higher value of $m_{\tg}\sim 2.4$ TeV with
concomitantly heavier $\tw_1$ and $\tz_{1,2}$, although all EW gauginos along with 
$\tst_1$, $\tst_2$ and $\tb_1$ are below 1 TeV. 
The ${\tb_2}$ and staus all have masses $\sim 1.3-1.4$ TeV range. 
For this point, $\tst_1$ decays exclusively via the three-body
mode $\tst_1\to bW\tz_1$, while other third generation quarks decay via
two-body decays. 

Point ES3, with a value of $m_0(1,2)\simeq 10$ TeV, again has all third
generation sfermions, as well as all charginos, neutralinos and Higgs
bosons essentially at or well below 1~TeV. 
The gluino mass is rather high: $\sim 2.1$ TeV. One characteristic
feature of this point is that, because of the near degeneracy of $\tst_1$
and $\tz_1$, not only the two-body decays but even the three body decay
$\tst_1 \to bW\tz_1$ is kinematically forbidden. In this case, $\tst_1$
will decay via $\tst_1 \to c\tz_1$ or $\tst_1 \to b\tz_1 f\bar{f}'$, though of
course the small available phase space will favour the former decay~\cite{andrew}.

Point ES4 has a high $m_0(1,2)=20$ TeV, a heavy gluino with a mass of 2.5 TeV, and
correspondingly heavy EW gauginos. However, $\tst_1$ and $\tb_1$
are very light with masses of 327 GeV and 291 GeV, while
$m(\tst_2)=793$ GeV. In this case, $\tb_1$ is the
LMP. 

Finally, point ES5 has been chosen to illustrate that staus can also be
very light. Here,  $\ttau_1$ is the LMP, with a mass of just 289 GeV,
while $m(\ttau_2)=380$~GeV. The lighter stop and the EW gauginos 
are in the sub-TeV range, while other third generation sfermions are 
1.1-1.2~TeV.

\begin{table}
\begin{center}
\begin{tabular}{lccccc}
\hline
parameter & ES1 & ES2 & ES3 & ES4 & ES5 \\
\hline
$m_0(1,2)\ [{\rm TeV}]$  & 10 & 12 & 10 & 20 & 6 \\
$m_0(3)\ [{\rm TeV}]$    & 2.4 & 1.5 & 1.1 & 3.3 & 0.3 \\
$m_{H_{d}}\ [{\rm TeV}]$& 1 & $-0.35\phantom{0}$ & $-0.6\phantom{0}$ & 1 & 0 \\
$m_{H_{u}}\ [{\rm TeV}]$& 2 & 1.2 & 0.5 & 1 & 0 \\
$m_{1/2}\ [{\rm TeV}]$   & 0.16 & 1 & 0.85 & 1 & 0.8 \\
$A_0$ [{\rm TeV}] & 0 & 0.9 & 1 & 0 & 0 \\
\hline
$\mu$       & 857 & 902 & 925 & 2329 & 855 \\
$m_{\tg}$   & 524 & 2446 & 2103 & 2526 & 1941 \\
$m_{\tu_L}\ [{\rm TeV}]$ & 10.0 & 12.1 & 10.1 & 20.1 & 6.2 \\
$m_{\tu_R}\ [{\rm TeV}]$ & 10.0 & 12.2 & 10.1 & 20.1 & 6.2 \\
$m_{\te_L}\ [{\rm TeV}]$ & 10.0 & 12.0 & 10.0 & 20.0 & 6.0 \\
$m_{\te_R}\ [{\rm TeV}]$ & 10.0 & 12.0 & 10.0 & 20.0 & 6.0 \\
\hline
$m_{\tst_1}$& 646 & 608 & 398 & 327 & 867 \\
$m_{\tst_2}$& 1049 & 948 & 770 & 793 & 1147 \\
$m_{\tb_1}$ & 1039 & 830 & 586 & {\bf 292} & 1083  \\
$m_{\tb_2}$ & 1711 & 1313 & 958 & 1885 & 1176.3\\
$m_{\ttau_1}$ & 2269 & 1341  & 944 & 2956 & {\bf 289} \\
$m_{\ttau_2}$ & 2299 & 1388  & 1008 & 3152 & 381 \\
\hline
$m_{\tw_1}$ & 139 & 815 & 708 & 881 & 658 \\
$m_{\tz_2}$ & 139 & 815 & 708 & 878 & 657 \\
$m_{\tz_1}$ & {\bf 69} & {\bf 441} & {\bf 372} & 452 & 347 \\
$m_A$       & 1022 & 450 & 398 & 2042 & 875 \\
$m_h$       & 110.7 & 118.3 & 117.3 & 119.9 & 117.8 \\
\hline
$\Omega_{\tz_1}h^2$ & 320 & 0.789 & $0.036$ & -- & --\\
\\
\hline
$\sigma\ [{\rm fb}]$ & $23.2\times 10^3$ & $157.8$ & $1618.0$ 
& $12.3\times 10^3$ & $51.4$ \\
$\tg\tg$             & 62.8\% & 0.02\% & 0.01\% & -- & 1.5\% \\
$EW-ino\ pairs$      & 36.6\% & 5.1\% & 0.8\% & 0.06\% & 35.3\% \\
$slep.\ pairs$       & -- & 0.02\% & 0.01\% & -- & 24.4\% \\
$\tst_1\bar{\tst}_1$ & 0.4\% & 78.3\% & 87.6\% & 28.2\% & 26.5\% \\
$\tst_2\bar{\tst}_2$ & 0.03\% & 4.5\% & 1.8\% & 0.3\% & 3.8\% \\
$\tb_1\bar{\tb}_1$   & 0.02\% & 11.5\% & 9.1\% & 71.4\% & 4.9\% \\
$\tb_2\bar{\tb}_2$   & -- & 0.4\% & 0.4\% & -- & 2.9\% \\
\hline

\hline
\end{tabular}
\caption{Masses and parameters in~GeV units
(unless otherwise noted) for five case study points in ESUSY
using Isajet 7.80 with $m_t=173.1$ GeV, $\tan\beta =10$ and $\mu>0$.
The LMP mass is denoted with bold numbers.
We also list the
total tree level sparticle production cross section
in fb at the LHC with a center of mass energy of 14 TeV.
}
\label{tab:bm}
\end{center}
\end{table}

Two of the ESUSY points with a neutralino as LMP (ES1 and ES2) yield a
thermal abundance of neutralinos far in excess of WMAP measured value
for the cold DM relic density, while point ES3 is in the top-squark
co-annihilation region and gives a thermal neutralino 
abundance of $\Omega_{\tz_1}h^2 = 0.036$.  
As mentioned previously, the points ES1
and ES2 can still be cosmologically viable if we invoke the 
PQWW solution to the strong CP problem, which necessitates the introduction
of an axion/axino supermultiplet into the theory. In this case, the
axino $\ta$, and not the neutralino, can be the 
LSP. 
 If the $\ta$ is the LSP, then the neutralinos will decay via
$\tz_1\to \ta\gamma$ with lifetime smaller than $\alt$ 0.1--1\,s, {\it i.e.}
before the onset of Big Bang nucleosynthesis (BBN), for a PQWW scale $f_a\alt
10^{12}$~GeV~\cite{ckr}. Since every
neutralino decays to an axino, the would be thermal relic density of 
neutralinos is reduced by a factor
$m_{\ta}/m_{\tz_1}$, and for small enough $m_{\ta}$ would be 
compatible with the relic density measurement.  
Thermal production of axinos in the
early universe can still proceed, but the relic abundance of axinos then
depends on the re-heat temperature $T_R$ after inflation. The dark
matter will then consist of an axino/axion admixture~\cite{bbs}, with relic axions
being produced via the vacuum mis-alignment mechanism.

In the cases where $\tst_1$, $\tb_1$ or $\ttau_1$ are LMPs, such 
scenarios are again allowed if, as before, we assume that the
axino is the LSP.  In these cases the sfermions dominant decay is
$\tf_1\to f \ta$, via loop-mediated processes; the competing three-body
decays, $\ttau_1 \to \tau\ta\gamma$~\cite{branden} or $\tq\to q\ta
g$~\cite{small} are argued to have a small branching fraction. The loop
calculation is complicated by the fact that the non-renormalizable
axino-bino-photon (and the axino-bino-Z and axino-gluino-gluon) coupling
enters the calculation. Nevertheless, these authors find that for
$m_{\ttau}=100$~GeV, the stau LMP lifetime ranges from about 0.01\,s to 10\,h,
for the PQWW scale $f_a$ in the range $5\times 10^9$\,--\,$5\times
10^{12}$~GeV, and scales inversely as the stau mass.  The corresponding
squark lifetime is shorter: about $2\times 10^{-6}$\,s for 500~GeV squarks
and 1~TeV gluinos with $f_a=10^{11}$~GeV.  If the LMP lifetime indeed
exceeds a few seconds, this could disrupt the successful predictions of
Big Bang nucleosynthesis~\cite{jedamzik}, though it appears that if $f_a< 10^{12}$~GeV,
a stau LMP is relatively safe~\cite{freitas}. For a general discussion of
charged LMPs, see Ref.~\cite{berger}.

\section{Phenomenology of Effective SUSY at the LHC}
\label{sec:lhc}

\begin{figure}[t]
\begin{center}
\epsfig{file=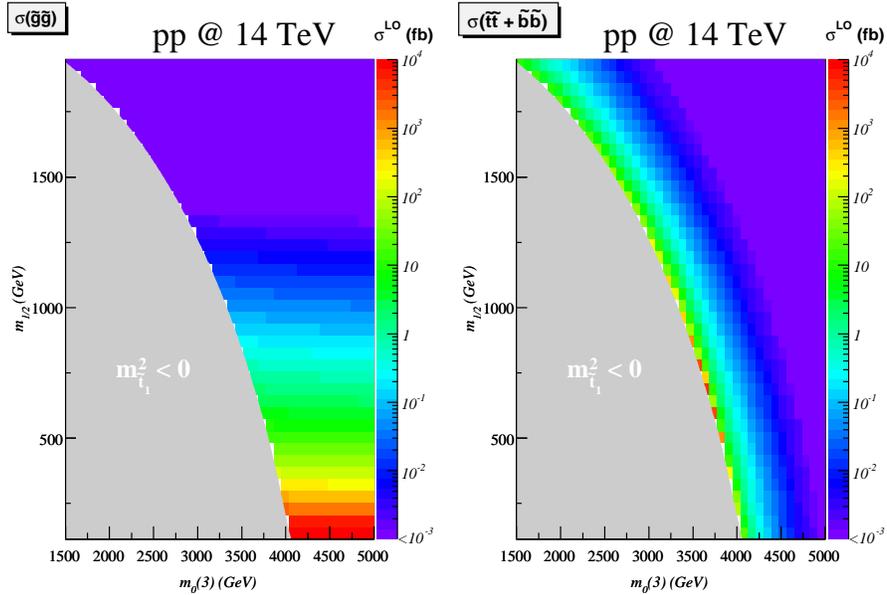,width=12cm}
\end{center}
\vspace*{-4mm}
\caption{\it 
{\it a}) Gluino pair production and {\it b}) third generation 
squark pair production cross sections at LHC with $\sqrt{s}=14$ TeV in the
$m_0(3)\ vs.\ m_{1/2}$ plane for $m_0(1,2)=20$ TeV, $m_{H_{u,d}}=1$ TeV
and $A_0=0$, $\tan\beta =10$ and $\mu >0$.
}\label{fig:xsecs14} 
\end{figure}

How will effective SUSY manifest itself at the LHC? To answer this, we
first show in Fig.~\ref{fig:xsecs14} total gluino and third
generation squark  pair production cross sections at a 14~TeV $pp$
collider
in the $m_0(3)\ vs.\
m_{1/2}$ plane displayed in Fig. \ref{fig:plane}, but now with
$m_0(1,2)=20$ TeV.  In frame {\it a}), the $pp\to\tg\tg X$ reaction
(where $X$ denotes assorted hadronic debris) reaches to over $10^4$ fb
for gluino masses as low as $m_{\tg}\alt 500$ GeV.  The gluino pair cross
section drops continuously as $m_{1/2}$, or alternatively  $m_{\tg}$,
increases.

The ESUSY region with third generation squarks having masses $\alt$TeV lies adjacent to 
the boundary of the gray tachyonic region. In frame {\it b}), we see the
$pp\to\tst_i\bar{\tst}_i$ plus $\tb_i\bar{\tb}_i$ (for $i=1-2$)
summed cross sections. These are typically in the 1-100 fb range
all along the tachyonic boundary, with portions reaching cross sections as
high as $10^3$ fb. Thus, the low $m_{1/2}$ region will consist
of mainly gluino pair production, where the main influence of light
third generation squarks will be upon the gluino branching fractions.
As we move to higher $m_{1/2}$ values, the gluino pair cross section will
diminish, and the total SUSY production cross section will be dominated
by pair production of third generation squarks.

\begin{figure}[th]{
\begin{center}
\includegraphics[height=8cm,width=5.6cm,angle=270]{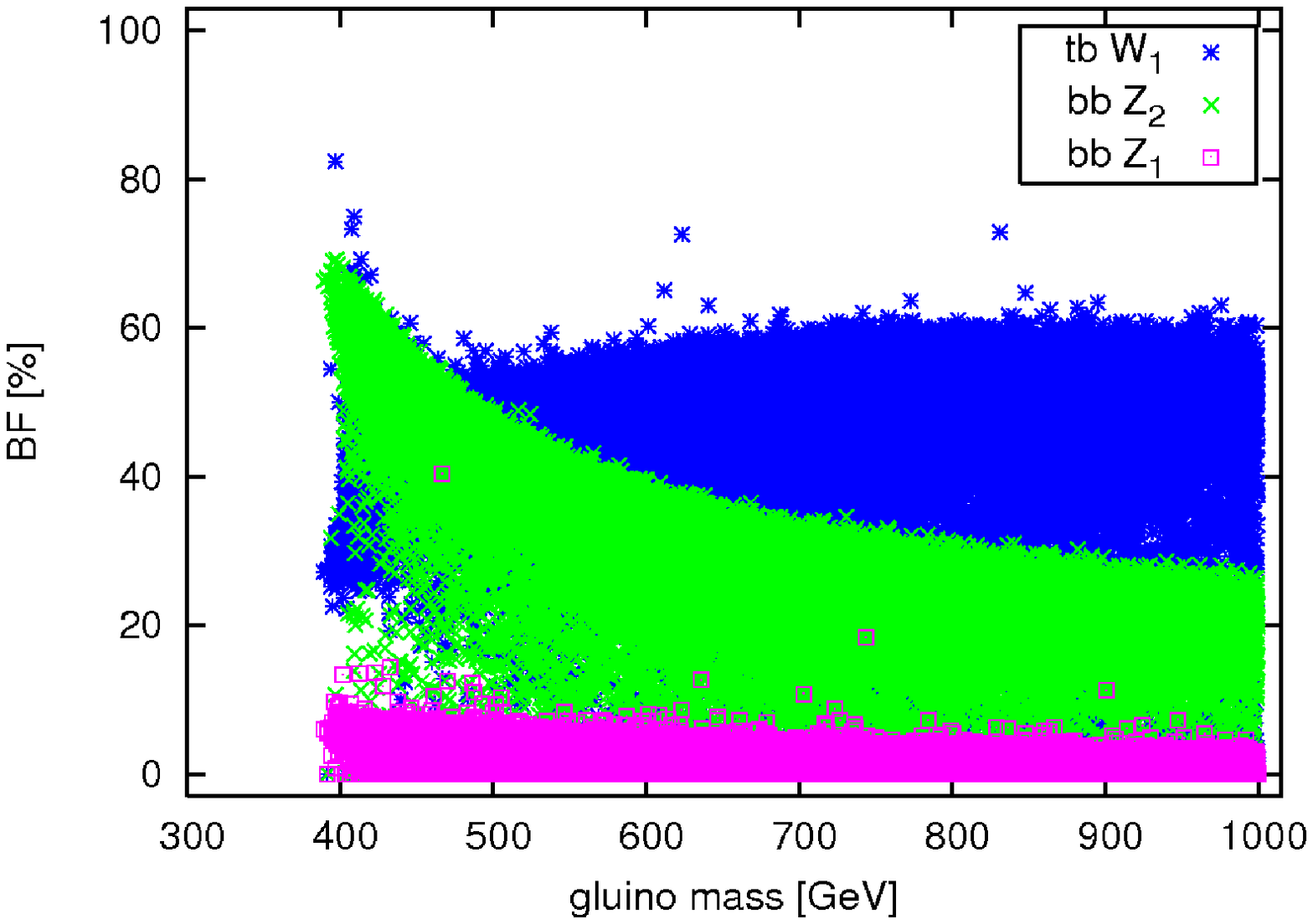}
\includegraphics[height=8cm,width=5.6cm,angle=270]{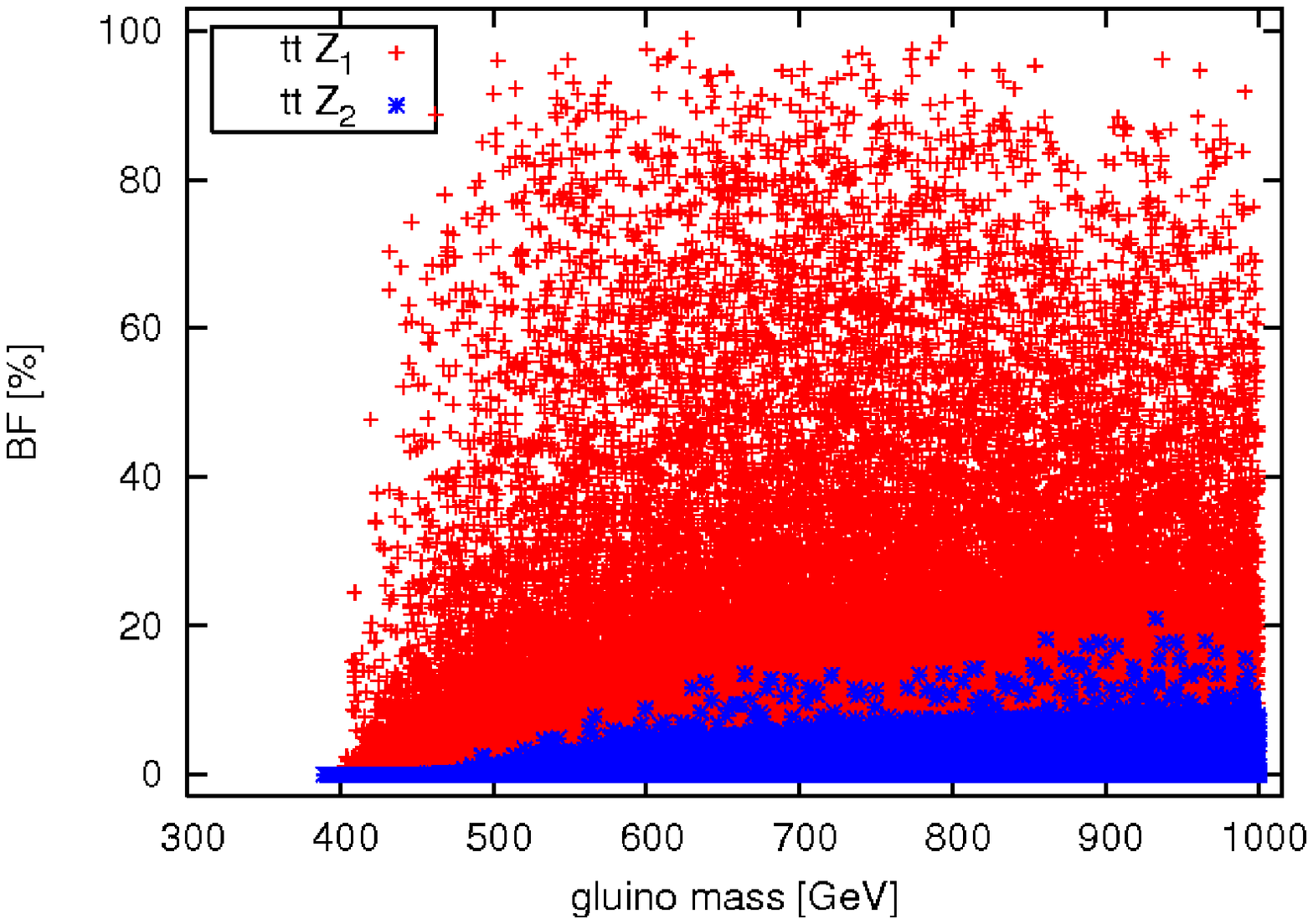}
\vspace{5pt}
\includegraphics[height=8cm,width=5.6cm,angle=270]{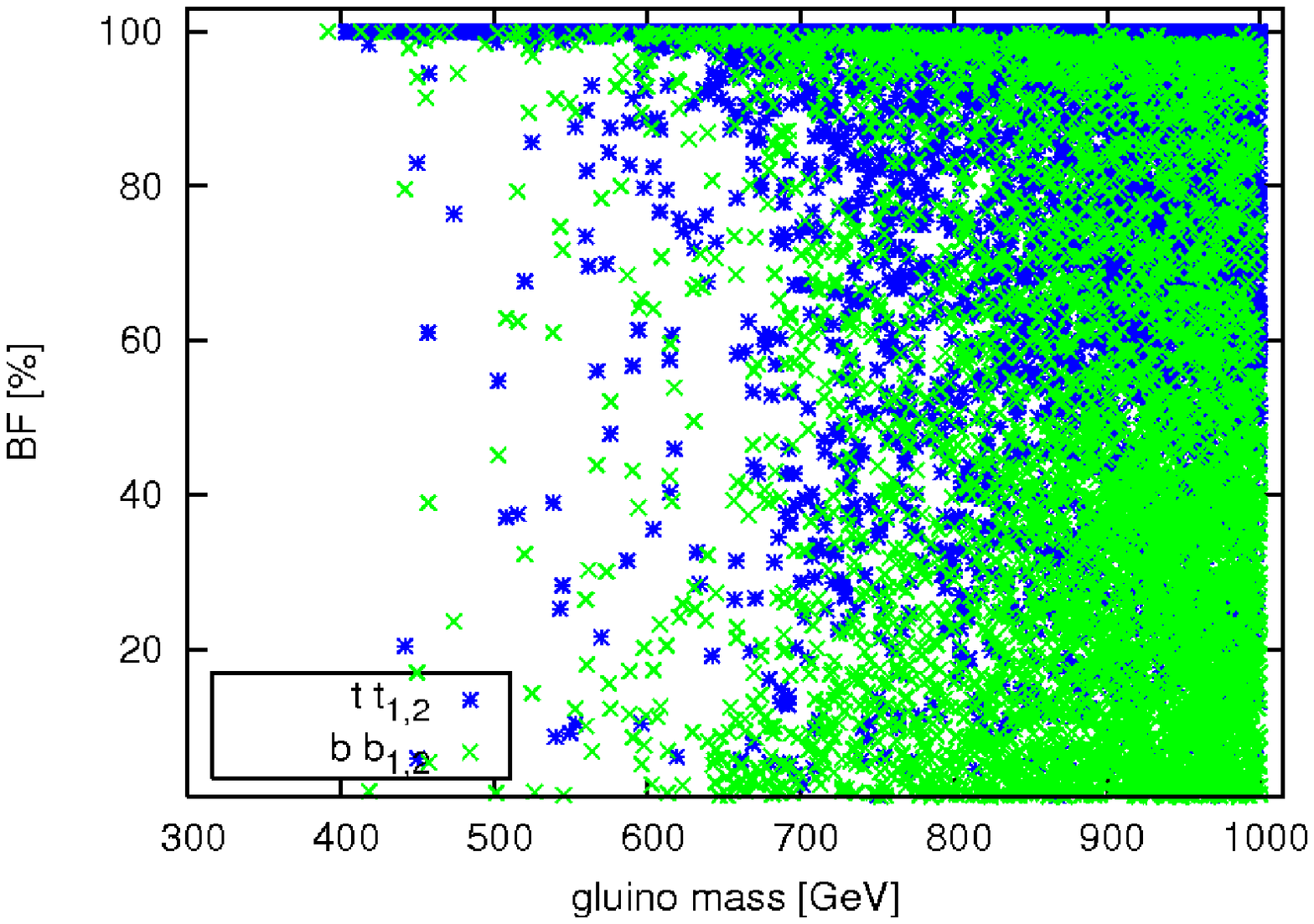}
\includegraphics[height=8cm,width=5.6cm,angle=270]{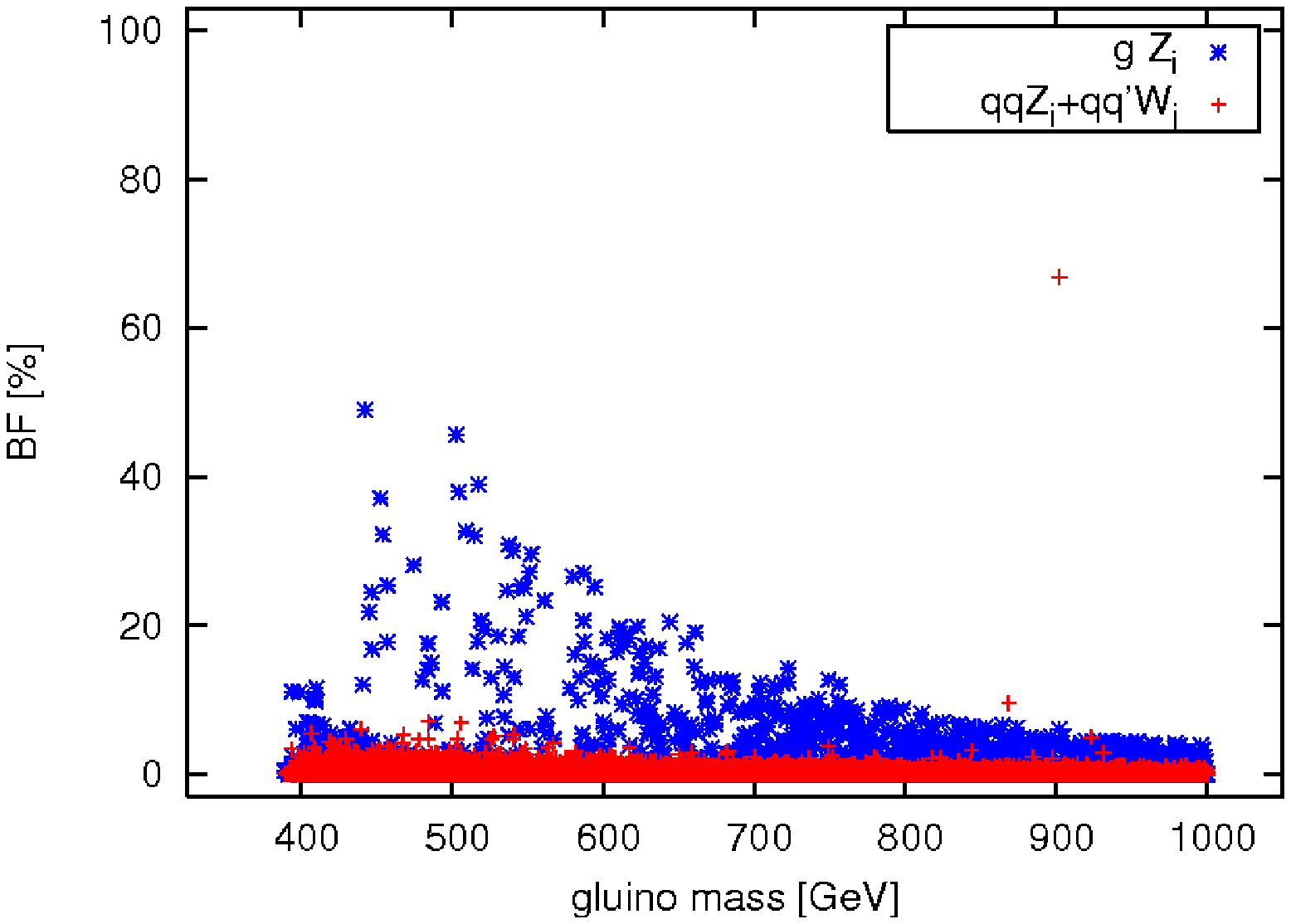}
\end{center}
\vspace*{-4mm}
\caption{\it Results for the branching fractions for gluino decays from
  MCMC scan points with $m_{\tg}\le 1$~TeV versus the gluino mass. The 
  upper two frames show the branching fractions for three-body decays of gluinos 
  to third generation quarks plus EW gauginos. The bottom-left 
  frame shows the branching fraction for two-body decays to third
  generation quarks and squarks, while the bottom-right frame illustrates that 
  the
  radiative decays $\tg\to g\tz_i$ can sometimes be substantial. All results
  are for the $L_{\widetilde M}= 1$ case with 5 TeV $< m_0(1,2)<$ 20 TeV.}
\label{fig:gluinodec}}
\end{figure}

Unless the gluino is very heavy we would expect that, especially after
selection cuts, LHC phenomenology would largely be determined by gluino
pair production. SUSY event topologies would thus be sensitive to gluino
decay patterns. In Fig.~\ref{fig:gluinodec}, we show the gluino
branching fractions for the MCMC scan points with $m_{\tg}\le 1$~TeV.
Of course, when gluino two-body decays are kinematically accessible,
these would dominate. We see from the figure that these are accessible
for relatively few points when $m_{\tg}< 500$~GeV. We will return to
this later when we discuss how one might distinguish the ESUSY model
with light gluinos from the model with $t$-$b$-$\tau$ Yukawa unification
where $m_{\tg}$ is bounded from above~\cite{yuk}. For the bulk of the
scanned points, the gluino decays via three body modes into top and
bottom quarks plus EW gauginos; corresponding decays to the first
two generations are suppressed because these squarks are very
heavy. Bottom-jet tagging will clearly provide an effective way for
enriching the ESUSY event sample at the LHC~\cite{btag,so10lhc1}. 
As in many SUSY models,
direct decays to the neutralino LMP never have a very large branching
fraction. The fact that gluino decays have large branching fractions to
$\tw_1,\,\tz_2$ and top quarks implies that the ESUSY gluino event sample
will include multi-lepton events from gluino decay cascades. Finally, we
see from the last frame that while the branching fraction for the
radiative gluino decays~\cite{radglu} $\tg\to g\tz_i$ is usually small, 
it can reach 40--50\%. 
We have checked that the neutralino in question is mostly $\tz_{3,4}$. 
This occurs when we have large splittings among third
generation squarks with stops being lighter than sbottoms, when a light
neutralino has a significant up-higgsino component and the decay $\tg\to
t\bar{t}\tz_1$ is kinematically forbidden.  The branching fraction for
decays to light quarks is very small for $m_0(1,2)>5$~TeV, although larger 
values are possible with the second $L_{\widetilde M}$ prior 
(case 2. in Sec.~\ref{ssec:scans}), which disfavours lower values 
of $m_0(1,2)$ but does not exclude them.

\subsection{Light gluinos with $\tz_1$ as LMP (ES1)}

The benchmark point ES1 features a rather light gluino with mass
$m_{\tg}=524$ GeV, correspondingly light charginos and neutralinos,
along with three sub-TeV third generation squarks. From
Table~\ref{tab:bm}, we see that total sparticle production cross
section, summed over all reactions, is 23.2~pb. Of this, 62.8\% comes
from $pp\to\tg\tg X$ production, while 36.6\% comes from EW gaugino 
production (sum of all chargino and neutralino production cross
sections).  Only a tiny fraction of the production cross section comes
from third generation squarks, so the rate for $\tst_i\bar{\tst}_i$ and
$\tb_i\bar{\tb}_i$ production (for $i=1,2$) is very small.

The gluino decays via three body modes as $\tg\to b\bar{b}\tz_2$ (37\%),
$\tg\to t\bar{t}\tz_1$ (19\%) and $\tg\to t\bar{b}\tw_1+ c.c$ (41\%).  Thus,
each gluino pair producton event will contain at least four $b$-jets,
plus other jets, isolated leptons (from top quarks and EW gauginos)
and $\eslt$. The $\tw_1$ decays via three-body modes $\tw_1\to\tz_1
f\bar{f}'$ in accord with $W^*$ propagator dominance ({\it e.g.}
$\tw_1\to\tz_1 e\bar{\nu}_e$ at 11\%).  The $\tz_2$ decays to
$b\bar{b}\tz_1$ 20\% of the time, while leptonic decays such as
$\tz_2\to\tz_1 e^+e^-$ occur at the 3\% level. In this case,
the $\tb_1$ is quite light and dominantly $\tb_L$,
which enhances the $\tz_2$ decay to $b$ quarks at the expense of first and
second generation fermions. Clean trilepton signals from $\tw_1\tz_2$
production may also be observable~\cite{trilep_lhc}.

\begin{figure}[t]
\begin{center}
\epsfig{file=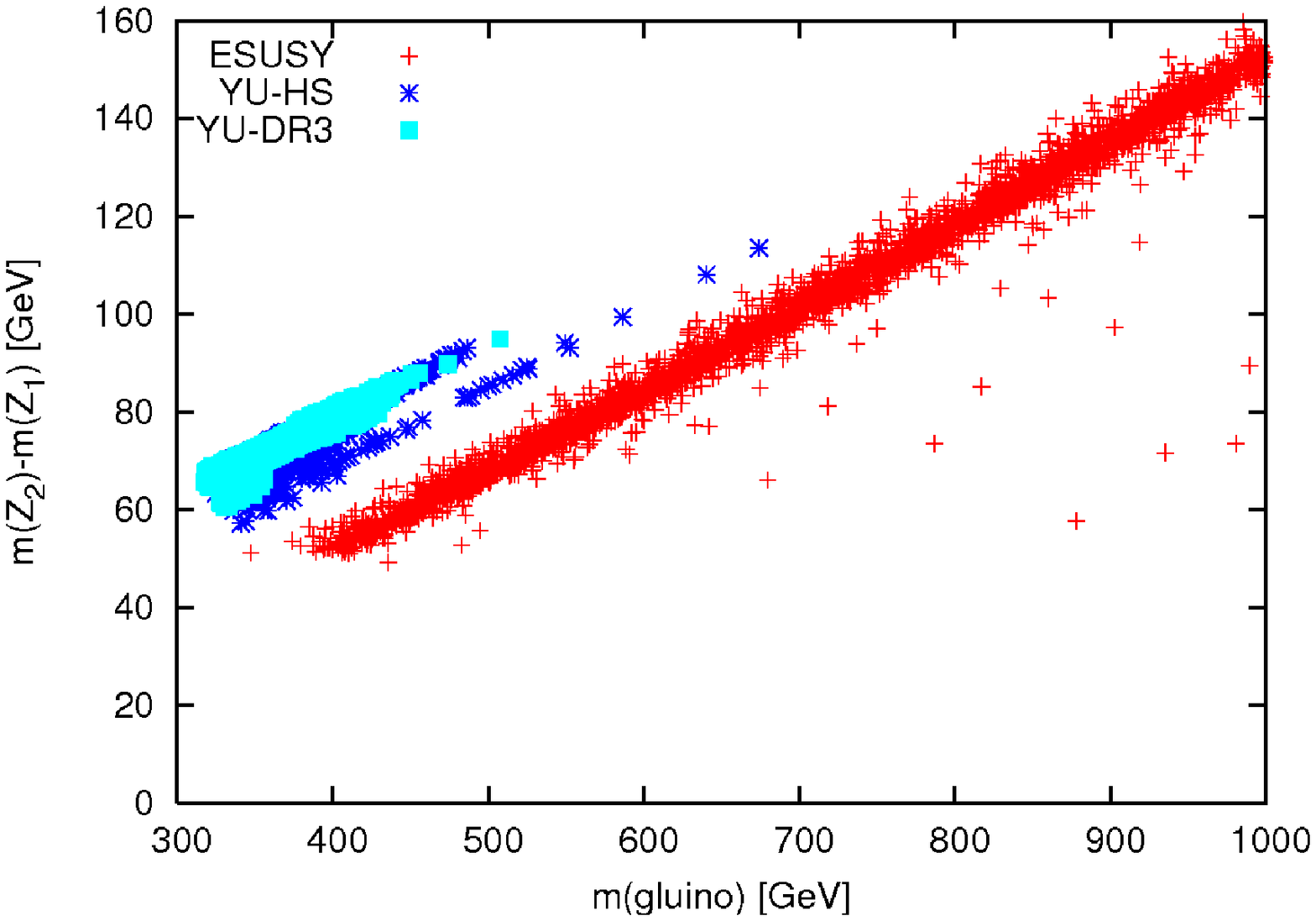,width=8cm,angle=270}
\end{center}
\vspace*{-4mm}
\caption{\it Scatter plot of the gluino mass versus the neutralino
  mass difference $m_{\tz_2}-m_{\tz_1}$ that can be obtained from the
  determination of  the
  dilepton mass edge at the LHC. We show results for the ESUSY model (red),
  and two different models with $t$--$b$--$\tau$ Yukawa unification (medium 
  and light blue) discussed in the text.
}\label{fig:diff} 
\end{figure}

The ES1 scenario will have much the same LHC phenomenology as the
Yukawa-unified SUSY with a 500 GeV gluino~\cite{so10lhc1}, which is also
dominated by gluino pair production followed by three-body gluino decays
into mainly $b$-quarks. A natural question to ask is whether it is
possible to distinguish these scenarios at the LHC.  One difference is
that Yukawa-unified SUSY requires a very large $A_0$ parameter with
$A_0\sim -2m_{16}$, where $m_{16}$ is the matter GUT scale scalar mass
parameter, which lies typically in the range 8--20~TeV. This large 
value of $A_0$ feeds into gaugino mass evolution at two-loops in the
RGEs, and suppresses $m_{\tg}$ with respect to $m_{\tz_1}$ and
$m_{\tz_2,\tw_1}$. 

To illustrate this, we show in Fig.~\ref{fig:diff} a scatter plot of 
$m_{\tg}$ versus $m_{\tz_2}-m_{\tz_1}$ obtained by scanning the
parameter space of (a)~the ESUSY model (red pluses) and (b)~the Yukawa-unified (YU)
model with ``just-so'' GUT scale Higgs soft mass splitting (HS, dark blue stars) 
and the YU model with full $D$-term splitting, right-hand neutrinos and 
3rd generation scalar splitting (the DR3 model~\cite{dr3}, light blue squares). 
Here, we require in YU models that $R$, the ratio of the largest to the smallest of the GUT scale
Yukawa couplings, is smaller than $1.05$.  While there is significant
overlap of the two classes of Yukawa-unified models, the ESUSY model
lies on a distinct band. We expect that the neutralino mass difference edge
will be rather well determined by the $m(\ell^+\ell^-)$ distribution. 
Then, even a crude measure of the gluino mass at the 
10--20\% level via $M_{eff}$~\cite{meff}
or total cross section~\cite{gabe} would suffice for this distinction. 
Also, in Ref.~\cite{so10lhc1}, it 
is shown that the mass difference $m_{\tg}-m_{\tz_2}$ might be extracted
via the kinematic edge in the $m(bb)$ distribution.\footnote{
We have also checked that a scatter plot of $m_{\tg}-m_{\tz_2}$ vs,
$m_{\tz_2}-m_{\tz_1}$ also serves to separate the ESUSY model from the
Yukawa-unified SUSY scenarios, but do not show it here for brevity.} 
If $m_{\tg}$ can be determined even
more precisely using $m_{T2}$~\cite{mt2}, its generalizations~\cite{mtgen}, 
or other methods that have recently been suggested, the
distinction between the models would be even sharper.

\subsection{Light top and bottom squarks, heavy gluinos and $\tz_1$ as LMP (ES2 \& ES3)}

For point ES2, since $m_{1/2}$ is large, the value of $m_{\tg}$ is also
large, and so $\tg\tg$ will not be produced at an appreciable rate at
LHC.  Instead, the total SUSY cross section of $\sigma_{SUSY}=157.8$~fb
is dominated by 78.3\% $\tst_1\bar{\tst}_1$ production, and 11.5\%
$\tb_1\bar{\tb}_1$ production. A few percent of $\tst_2\bar{\tst}_2$
also contribute. The $\tst_1\to bW\tz_1$ decay occurs at nearly 100\%
branching fraction, so the $\tst_1\bar{\tst}_1$ production results in a
$b\bar{b}W^+W^- +\eslt$ final state. This should be separable from
$t\bar{t}$ background if large $\eslt$ or $M_{eff}$ is required. 
Another possible
background is $Zt\bar{t}$ production. Because gluinos are heavy, SUSY
contamination to the signal~\cite{kadala} is not an issue. The $\tb_1\to
W\tst_1$ decays at nearly 100\% branching fraction, so this component of the
production cross section will result in a $b\bar{b}W^+W^-W^+W^- +\eslt$
final state.

The case of ES3 has four sub-TeV third generation squarks plus
tau-sleptons near a TeV or below. The total production cross section of
$\sigma_{SUSY}=1618$ fb comes from 87.6\% $\tst_1\bar{\tst}_1$
production, 9.1\% $\tb_1\bar{\tb}_1$ production plus a few percent of
heavier top and bottom squark pairs.  The visible decay products from
$\tst_1$ decay will be soft since there is only a 26 GeV mass gap
between the $\tst_1$ and the $\tz_1$; the visible decay products of $\tst_1$ 
may not reliably detectable unless they are boosted. 
The $\tb_1\to W^-\tst_1$ at nearly 100\%
branching fraction, so from $\tb_1\bar{\tb}_1$ production we expect a
$W^+W^- +\eslt$ final state, accompanied by soft charm jets or other
soft hadronic debris. The possible
background comes from $W^+W^-Z$ production.

Before turning to the next benchmark case we remark that, although we
have not examined an example of such a benchmark point, it has been
shown~\cite{nojpop} that if the gluinos and stops both have sub-TeV
masses and that the decays $\tg\to t\tst_1 \to tb\tw_1$ and/or $\tg \to
b\tb_1 \to bt\tw_1$ are kinematically accessible and occur with large
branching fractions, a partial reconstruction of the event using
techniques to tag the secondary top quark may be possible.
Moreover, in~\cite{Kraml:2005kb} it has been shown that if 
$\tg\to t\tst_1 \to tc\tz_1$ dominates, the signature of 2 b-jets 
plus 2 same-sign leptons plus additional jets and missing energy 
is an excellent discovery channel for gluino masses up to about 900 GeV.

\subsection{Bottom or top squarks as LMP: 
the case of heavy quasi-stable colored particles (ES4)}

Benchmark point ES4 illustrates an ESUSY model with 
$\tst_1$ and $\tb_1$ lighter than $\tz_1$, 
and in fact the bottom squark $\tb_1$ is the
LMP. This is viable if the axino is the true  LSP
so that $\tb_1\to b\ta$, and the dark matter
consists of an axion/axino admixture. 
For case ES4, with a PQ breaking scale of order $f_a \sim 10^{11}$ GeV, we find 
a bottom squark lifetime of $\sim 10^{-6}$ sec, {\it i.e.} the
$\tb_1$ is stable as far as collider searches go, but it decays well 
before the start of Big Bang nucleosynthesis.
Previous searches for quasi-stable squark hadrons $Q$ at ALEPH find
that $m_Q>95\ (92)$ GeV for up-type (down-type) squark hadrons~\cite{aleph}.
Searches by CDF evidently require $m_Q\agt 241.5$ GeV~\cite{cdf,raklev}.
At the LHC, for case ES4, sparticles will be produced with total
cross section $\sigma_{SUSY}\simeq 1.2\times 10^4$ fb, of which
71.4\% is $\tb_1\bar{\tb}_1$ production, and 28.2\% is $\tst_1\bar{\tst}_1$
production. The $\tst_1\to \tb_1 W$ branching fraction is essentially 100\%,
and so augments the $\tb_1$ production rates. 
The $\tb_1$ is a quasi-stable colored particle, and will hadronize
to form an $R$-meson or baryon $B^{\pm ,0}=\tb_1 \bar{q}$ or $\tb_1 qq'$,
where $q$ is usually $u$ or $d$. The properties of squark hadrons
have been reviewed in~\cite{fairbairn,raklev}. The heavy $B$ hadron
will be produced and propagate through the detector with minimal
energy loss due to hadronic interactions. It can be of charge $\pm 1$ or $0$,
and in fact as it propagates, it has charge exchange reactions with
nuclei in the detector material, so that
the path of the $B$-hadron can thus be intermittently charged or
neutral~\cite{dt,bcg}.

Stable squark hadrons can be detected as muon-like events, albeit
with the possibility of intermittently appearing and disappearing tracks.
Detection can be made using $dE/dx$ measurments in the tracking system, or
time-of-flight (ToF) measurements in the muon system. Measuring both the $B$ hadron momentum
and velocity should allow a $B$ mass measurement to about 1-2 GeV~\cite{raklev}.
The reach of LHC for quasi-stable squark hadrons has been estimated 
in Ref.~\cite{raklev}. Using those results, we estimate the 
LHC reach with $\sqrt{s}=7$ TeV and 1 fb$^{-1}$ of integrated luminosity
to be up to $m_{\tb_1} \sim 315$ GeV, which would already test point
ES4! The LHC reach with $\sqrt{s}=14$ TeV and 100 fb$^{-1}$ is 
estimated as $m_{\tb_1}\sim 800$ GeV, from just $\tb_1\bar{\tb}_1$
production alone.  Additional contributions to $\tb_1$ 
production from cascade decays would increase the reach, and also 
make clear that $\tb_1$-pair production is not the only new physics
process occuring at the LHC.

We note also that many cases with a $\tst_1$ LMP can also be generated
in ESUSY. In these cases, with a quasi-stable top-squark hadron, 
the lifetime and reach discussion is qualitatively 
similar to that given above.

\subsection{Staus as LMP: 
the case of heavy quasi-stable charged particles (ES5)}

For ESUSY benchmark point ES5, we have $\ttau_1$ as LMP. As with point ES4,
we may again invoke a lighter axino
to escape constraints on charged stable exotics and 
also mixed axion/axino CDM. 
In this case, the stau decays through loop diagrams via 
$\ttau_1\to \tau\ta$ with a lifetime typically of order 1 sec for
$f_a\sim 10^{11}$~GeV, and does not significantly disrupt BBN~\cite{freitas}. 
Searches for quasi-stable charged particles at OPAL require
$m_{\ttau_1}\agt 98.5$ GeV~\cite{opal}.

The quasi-stable stau will propagate slowly though LHC detectors much
like a heavy muon, and leave a highly-ionizing track~\cite{dt}.  The
$dE/dx$, ToF and track bending measurements should allow for momentum
and velocity, and hence mass determinations.

For case ES5, the total SUSY production cross section is
$\sigma_{SUSY}\simeq 51.4$ fb. Of this, just about a quarter comes from
stau pair production. However, since all produced sparticles now cascade
down into the stau LMP, the total $\ttau_1$ production is augmented, in
this case, by a factor four. For instance, $\tst_1\bar{\tst}_1$ is
produced at 26.5\% rate, and $\tst_1\to \tz_1 t$ at 53.3\%, $\tst_1\to t\tz_2$ 
at 7.4\% and $\tst_1\to b\tw_1$ at 39.3\%. Then, $\tz_1\to
\tau^\pm\ttau_1^\mp$ at 100\% branching fraction. Also, $\tw_1\to
\ttau_1\nu_\tau$ at 100\%, and $\tz_2\to \tau^\pm\ttau_2^\mp$ at 42\%
and $\tnu_\tau\bar{\nu}_\tau +c.c$ at 49\%, while $\tnu_\tau\to
\tz_1\nu_\tau$, followed by further $\tz_1$ decay. Thus, we expect the
cascade decay events to each contain at least two $\ttau_1$ tracks,
frequently in company of an assortment of $\tau$-jets, together with $b$
quarks and leptons from $W$-bosons. If the energy of the LMP (as
measured from the velocity and momentum of the track) can be included,
$\eslt$ in such SUSY events arises only from neutrinos, and so is not
especially large.

The LHC reach for quasi-stable staus has been estimated in~\cite{raklev}
in the case where only stau pair production contributes. The reach for
LHC at $\sqrt{s}=14$ TeV with 100 fb$^{-1}$ of integrated luminosity is
found to be $m_{\tau_1}\sim 250$ GeV.  This is conservative for the
present case, since $\ttau_1$ production will be augmented by a factor
of four from production via cascade decays of heavier
sparticles. Moreover, as in the ES4 case, detection of more complicated
events other than just stau pair production will bring home the richness
of the new physics being detected.

\section{Summary and Conclusions}
\label{sec:conclude}

Effective SUSY has been suggested as a model for ameliorating the
flavour and $CP$ problems that are endemic to generic SUSY models, while
maintaining naturalness in the EWSB sector. The general idea is that
SUSY states that have large couplings to the Higgs sector --- third
generation sfermions and EW gauginos --- have masses at or below the TeV
scale, while first and second generation sparticles that directly affect
the most stringently constrained flavour-changing processes are
heavy. The typical ESUSY spectrum consists of multi-TeV first and second
generation sfermions along with EW gauginos, Higgs bosons and third
generation sfermions at or below the TeV scale; gluinos may be as light
as 450~GeV or as heavy as several TeV.

In this paper, we have examined the LHC phenomenology of ESUSY
models. Toward this end, we set up the ESUSY model with parameters
defined at the GUT scale in Sec.~\ref{sec:intro}. For simplicity
of incorporting flavour constraints without greatly impacting LHC
physics, we have assumed GUT scale degeneracy of the SSB mass parameters of
the first two generations, but allowed the corresponding parameters for
the third generation and Higgs scalars to be independent.

We have delineated the viable parameter space in
Sec.~\ref{sec:pspace}. In Sec.~\ref{ssec:scans}, we have described the
MCMC setup that was used for finding the portion of the entire parameter
space consistent with various low energy constraints as well as lower
limits on sparticle and Higgs boson masses along with theory priors that
were used to obtain ESUSY spectra. Our results for the favoured ranges
of the input parameters, assuming that the LMP is a neutralino (this is
true for the bulk of the points), are shown by the posterior
probability distributions in Fig.~\ref{fig:mcmc1}. The corresponding
distributions for sparticle masses, illustrated in Fig.~\ref{fig:mcmc2},
indeed show the expected qualitative features of the ESUSY spectrum:
$\alt$TeV third generation and EW gaugino masses, 10-20~TeV first/second
generation squarks (and sleptons, that we did not show), and gluinos
ranging from 0.5-4~TeV.  The posterior probability distributions for
various low energy observables and the thermal neutralino relic density
are shown in Fig.~\ref{fig:mcmc3}. While the former generally lie close
to their SM values (because of the nature of the ESUSY spectra), the
neutralino relic density $\Omega h^2$ ranges from below 0.01 to $10^3$
with a peak value around $1-10$.

We note that models with large values of $\Omega_{\tz_1} h^2$ are
phenomenologically every bit as viable as models with a thermal
neutralino LSP provided the LMP neutralino is not the true LSP, but decays
into the LSP before the advent of Big Bang nucleosynthesis.  The axino
of the axion-axino multiplet which is present in SUSY models with the PQWW 
solution to the strong $CP$ problem is an excellent example of an LSP
that is not the LMP, and indeed models where the measured CDM is
comprised of axions and axinos have been studied in the
literature. Moreover, if the axino is the LSP, models where the LMP is
charged or coloured are also allowed provided again that the LMP decays
before nucleosynthesis. 

Our main results on the LHC phenomenology of ESUSY models were presented in
Sec.~\ref{sec:lhc}. The most important sparticle production mechansims
at the LHC are gluino pair production for values of $m_{1/2}\alt
700$~GeV, and third generation squark pair production.  A very high $b$-jet
multiplicity is the hallmark of multi-jet+multilepton+$\eslt$ events
within the  ESUSY framework. Gluinos mostly decay via (real or virtual) third
generation squarks (see Fig.~\ref{fig:gluinodec}) to third generation
quarks and EW gauginos. While chargino decay branching fractions
tend to follow those of the $W$-boson, the branching fraction for
neutralinos to $b$ quarks is often enhanced in these scenarios. Third
generation squarks frequently also have $b$-quarks as their decay
products. Finally, ESUSY events are also often rich in $t$ quarks whose
decays contribute to the multi-lepton component of the signal. 

We have discussed the phenomenology of five ESUSY benchmark points introduced
in Sec.~\ref{ssec:BM}. Of these, just the point ES3 has a thermal
neutralino relic density in accord with the measured cold DM relic
density, although here the $\tz_1$ would just make up about $1/3$ 
of the DM. 
For the other points, the LMP (which need not even be a
neutralino) must decay into the true LSP (which might be the axino)
plus SM particles. If the LMP is a neutralino, this decay typically
occurs well after the neutralino has passed through the LHC detectors,
and the LHC phenomenology is essentially the same as for a stable
neutralino. The LMP may also be a coloured squark or an electrically
charged stau, as illustrated by the last two benchmark points that we
discuss. Assuming again that the axino is the LSP, the LMP is
quasi-stable and typically traverses the detector before it decays.
In these cases, the presence of a (possibly intermittent,
in the case of squark LMP) track of a slowly moving particle (whose
velocity is obtained through timing information), rather than $\eslt$,
will be the hallmark of SUSY events. 

To conclude, effective supersymmetry with third generation
squarks and EW gauginos at or below the TeV scale, but 10--20~TeV 
first/second generation
sfermions ameliorates the flavour and $CP$ problems that plague many
SUSY models. We have examined the LHC phenomenology of
relic-density-consistent ESUSY scenarios and shown that it may differ
qualitatively from that in the most frequently examined mSUGRA
scenario. The ESUSY picture may also be tested in low energy
measurements. In particular, if the branching fraction for $B\to
\tau\nu$ decay, and especially  the muon magnetic moment
 show significant deviation from their SM values, the ESUSY picture
 would be strongly disfavoured.

\section*{Acknowledgements}

We thank A. Pukhov for his help in implementing the {\tt Isajet} -- {\tt micrOMEGAs} interface for SUSY scenarios with non-universal scalars.  XT thanks the UW IceCube collaboration for making his visit
to the University of Wisconsin where this work was done,
possible. We also thank the Galileo Galilei Institute (GGI), Florence,
Italy for hospitality at the time of the Workshop on Dark Matter: Its origin,
nature and prospects for detection during which many of the ideas presented
in this paper came together. 
This work was funded in part by the US Department of Energy,
and by the French ANR project 
ToolsDMColl, BLAN07-2-194882.


\end{document}